\documentclass{article}

\usepackage{arxiv}

\usepackage[utf8]{inputenc} 
\usepackage[T1]{fontenc}    
\usepackage{hyperref}       
\usepackage{url}            
\usepackage{booktabs}       
\usepackage{amsfonts}       
\usepackage{nicefrac}       
\usepackage{microtype}      
\usepackage{lipsum}		
\usepackage{graphicx}
\usepackage{natbib}
\usepackage{doi}
\usepackage{graphicx}
\usepackage{newtxtext}
\usepackage{newtxmath}
\usepackage{natbib}
\usepackage{multirow}
\usepackage{booktabs}
\usepackage{lipsum}

\usepackage{hyperref}
\hypersetup{
    colorlinks = true,
    urlcolor   = blue,
    citecolor  = black,
}

\newcommand{\RomanNumeralCaps}[1]

\title{Computational and reduced-order modelling of elastic wave-driven impulse enhancement in pulsed jets through passively flexible nozzles}

\date{} 					

\author{ 
    \href{}{Daehyun Choi}\thanks{Equal contributor}  \\
	School of Chemical and Biomolecular Engineering \\
    Georgia Institute of Technology \\
    Atlanta, GA, USA  \\
    \And
    \href{}{Paras Singh}\thanks{Equal contributor} \\
    Daniel Guggenheim School of Aerospace Engineering \\
    Georgia Institute of Technology \\
    Atlanta, GA, USA \\
	\And
	\href{}{Saad Bhamla}  \\
	BioFrontiers Institute, Department of Chemical and Biological Engineering \\
    University of Colorado Boulder, 
    Colorado, USA  \\
	\AND
	\href{https://orcid.org/0000-0001-5166-9968}{Chandan Bose}\thanks{Corresponding author}\\
	Aerospace Engineering, 
    School of Metallurgy and Materials\\
    University of Birmingham, 
    Birmingham, UK \\
	\texttt{c.bose@bham.ac.uk} \\
}

\renewcommand{\shorttitle}{Elastic wave-driven impulse enhancement in pulsed jets through passively flexible nozzles}

\hypersetup{
pdftitle={Computational and reduced-order modelling of elastic wave-driven impulse enhancement in pulsed jets through passively flexible nozzles},
pdfauthor={},
pdfkeywords={},
}

\title{Computational and reduced-order modelling of elastic wave-driven impulse enhancement in pulsed jets through passively flexible nozzles}

\begin{document}
\maketitle

\makeatletter
\@ifundefined{linenomathpatch}{%
  \@ifundefined{patchBothAmsMathEnvironmentsForLineno}{}{%
    \patchBothAmsMathEnvironmentsForLineno{equation}%
    \patchBothAmsMathEnvironmentsForLineno{align}}%
}{%
  \linenomathpatch{equation}%
  \linenomathpatch{align}%
  \linenomathpatch{gather}%
  \linenomathpatch{subequations}%
}%
\makeatother
\vspace{-1cm}
\begin{abstract}
This study investigates elastic wave propagation and energy exchange in passively deforming cylindrical nozzles through three-dimensional, two-way fluid--structure interaction simulations. Flexible nozzles with varying stiffness ($Eh = 75 - 500~\mathrm{N\,m^{-1}}$, where $E$ and $h$ are Young's modulus and nozzle thickness, respectively) are subjected to a pulsatile jet inflow \textcolor{black}{in the low Reynolds number regime ($Re \sim 4400$)}. Increasing nozzle flexibility reduces the deformation-wave speed in accordance with Moens--Korteweg scaling, thereby prolonging the nozzle expansion phase. This delayed expansion enhances jet entrainment and elastic energy storage while suppressing early shear-layer roll-up and vortex formation. During nozzle contraction, the stored elastic energy is released, enhancing jet acceleration and vortex formation. For the most flexible nozzle, the primary vortex-ring circulation increases by around 52\%, the vortex convection distance by 9\%, and the peak outlet kinetic energy flux by a factor of 4.6  compared with a rigid nozzle. These effects collectively yield a 62\% increase in total hydrodynamic impulse. \textcolor{black}{Additionally, a reduced-order model is developed that condenses the coupled fluid--structure response into a lumped store-and-release oscillator, derived as a single-mode projection of the inviscid one-dimensional wave equation and closed at the exit by two terms: (i) an inertial end correction that adds the external fluid column of length $L_e=R$ accelerating with the jet, and (ii) a vortex-radiation damping, active only during ejection, fixed by the momentum theorem for the discharged jet. This damping reproduces the post-overshoot velocity decay that undamped closures miss. The single-mode model reproduces the simulated resonance frequency to within $\sim6\%$ and the momentum impulse to within $4\%$ across $Eh=75$--$500~\mathrm{N\,m^{-1}}$, and recovers the store-and-release energy histories, the outlet kinetic-energy flux to within $6\%$ for the three stiffer nozzles and $14\%$ for the most compliant, giving a compact and predictive description whose assumptions are made explicit.}
\end{abstract}


\section{Introduction}
\label{sec:intro}
Pulsed jets, produced by the ejection of a finite volume of high-momentum fluid into a surrounding medium, are key to propulsion and fluid transport\textcolor{black}{, in combustion and process mixing \citep{li2024jet}, pulsating impingement cooling \citep{atofarati2024pulsating}, synthetic-jet flow control \citep{glezer2002synthetic,chiatto2015modelling}, pulsed-jet thrusters and underwater robots \citep{gao2020development,bujard2021resonant,choi2026squid}, and cardiovascular valve flows \citep{sun1995transmitral,vandevosse2011pulse}}. The pulsatile flows ejected from nozzles are dominated by the formation and evolution of vortex rings, which generate thrust through the combined effects of fluid momentum and pressure impulse at the nozzle or body exit \citep{krueger2003significance,krieg2015pressure}. In rigid nozzles, the generated impulse, composed of jet momentum and nozzle exit pressure contribution, \textcolor{black}{is set by the jet history the generator (piston or impeller) imposes through the velocity program \citep{rosenfeld1998circulation,shusser2006effect}, the discharged volume \citep{gharib1998universal,gao2020development}, and the nozzle geometry, fixed and non-circular \citep{ofarrell2014pinchoff,krieg2020vortex} or actively varied during ejection \citep{dabiri2005starting}.}

\textcolor{black}{Nature adopted pulsed-jet propulsion long before engineering did, cephalopods, salps and jellyfish swimming by expelling discrete slugs of fluid at high propulsive efficiency \citep{gosline1985jet,costello2021hydrodynamics,dabiri2009optimal,santoriello2026universality}, the airways ejecting multiphase jets in coughing and sneezing \citep{bourouiba2014violent}, compliant ducts voiding body fluids, the ureter in the peristaltic transport of urine \citep{griffiths1989flow,vahidi2012biomechanical}, the lactiferous duct in milk ejection \citep{ramsay2004ultrasound,ramsay2005anatomy}, the venom-gland duct in envenomation \citep{young2001venom,young2001mechanics}, the elephant proboscis (the trunk) drawing and expelling air and water \citep{schulz2021suction}, bivalves passing their feeding and respiratory currents through paired inhalant and exhalant siphons \citep{riisgard2011exhalant,uslu2016mytilus}, the archerfish varying the aperture of its mouth as it fires a water jet at prey \citep{gerullis2014archerfish}, the bombardier beetle pulsing its defensive spray through a cuticular valve \citep{arndt2015mechanistic}, the velvet worm ejecting its defensive slime through oscillating oral papillae \citep{concha2015oscillation}, and sap-feeding insects flicking excreted droplets clear of the body \citep{challita2023superpropulsion,ha2026developmental,challita2024fluid}. Each of these conduits is soft tissue, deforming under the pressure of the flow it carries \citep{wang2022fluid,demirer2022hydrodynamic,kim2025eversion}.}

\textcolor{black}{In a flexible nozzle, the wall deforms with the unsteady jet, and this coupling sets both the generated impulse and the propulsive efficiency \citep{dabiri2009optimal,wang2022fluid}, and how much the coupling adds to the thrust, and which flexibility maximises the gain, are the focus of recent work \citep{das2018unsteady,choi2022flow,choi2024mechanism,mitchell2025formation,cruciani2026global,santoriello2026flexibility,santoriello2026universality}: \citet{das2018unsteady} quantified the benefit in a two-dimensional channel with flexible flaps at the exit, introducing the non-dimensional flexural rigidity $EI^{*}=EI_{\mathrm{eq}}/(1/2\,\rho U_p^{2}L_f^{2}d)$ (with $EI_{\mathrm{eq}}$ the equivalent flap flexural rigidity, $L_f$ the flap length, $U_p$ the piston speed and $d$ the channel exit width) as the parameter that governs the flap response and showing that the flaps roughly double the jet impulse, with the extension from a planar flap pair to an axisymmetric nozzle left open. \citet{choi2022flow} gave a three-dimensional account for a continuous jet, deriving a governing non-dimensional wave speed $\hat c=\sqrt{Eh\,T_{\mathrm{acc}}^{2}/(\rho_f D L^{2})}$, the Moens--Korteweg wall wave speed $\sqrt{Eh/(\rho_f D)}$ measured against the nozzle length $L$ and the jet acceleration time $T_{\mathrm{acc}}$ (with $Eh$ the membrane stiffness, $\rho_f$ the fluid density and $D$ the nozzle diameter). At the optimum $\hat c\simeq3$ the wall stores elastic energy while the nozzle expands and returns it to the flow as the nozzle contracts, the release falling in phase with the jet, and the hydrodynamic impulse rises by about $100\%$ over a rigid nozzle. For a pulsed jet, \citet{choi2024mechanism} showed that nozzle collapse during deceleration suppresses the internal negative pressure and expels additional fluid, increasing the impulse by about $300\%$, and confirmed that the same optimum, $\hat c\simeq3$, still applies. Most recently, \citet{mitchell2025formation} showed at $Re=1000$, from planar particle image velocimetry of a syringe-pump-driven jet through silicone nozzles in a water tank, that this optimum is a resonance, the impulse peaking at up to about $180\%$ above the rigid nozzle when the jet-acceleration time equals one quarter of the damped natural period, $t_{\mathrm{acc}}=1/(4f_d)$ with $f_d$ the nozzle damped natural frequency, and forms multiple vortex rings. \citet{santoriello2026flexibility}, in three-dimensional direct numerical simulations with an immersed-boundary nozzle, attributed the thrust enhancement to a standing-wave, store-and-release response that raises the hydrodynamic impulse by up to about $240\%$, and derived a closed-form optimum, $K^{\mathrm{opt}}=16\,(L/T_p)^2$ (with $T_p$ the pulse duration).}

\textcolor{black}{Interest in flexible nozzles has grown, but reduced-order models for the compliant case remain scarce, even though they would benefit robotic and cardiovascular medical applications and clarify the underlying physics. Meanwhile, reduced-order models of pulsed-jet formation have been developed for rigid tubes.} \textcolor{black}{The classical slug model estimates the ejected circulation, impulse, and kinetic energy from the nozzle exit-velocity profile and forms the basis of the universal formation-number criterion for vortex-ring pinch-off \citep{gharib1998universal}. However, by assuming a spatially uniform exit velocity and zero pressure at the nozzle exit, the model underestimates the ejected circulation and impulse, as the over-pressure generated during jet initiation can contribute up to 42\% of the total impulse \citep{krueger2003significance,krieg2013modelling}. Incorporating a potential-flow correction for the nozzle-exit over-pressure reduces the circulation prediction error to approximately 6\% for a tube nozzle, compared with errors exceeding 20\% for the uncorrected slug model \citep{krueger2005overpressure}.}

\textcolor{black}{These closures all describe a rigid nozzle exit, whereas a compliant nozzle introduces wall motion and the elastic wave dynamics of a deforming conduit.} \textcolor{black}{One-dimensional models of pulsatile flow in compliant tubes, developed to describe arterial pulse-wave propagation and relate non-invasively measured pressure and flow waveforms to cardiovascular physiology \citep{hughes1973onedimensional,li2004dynamics,vandevosse2011pulse,mynard2015onedimensional,aghilinejad2025power}, capture the wave dynamics that wall elasticity produces in fluid-filled compliant conduits. These models close the downstream end in one of two ways. The first terminates the domain with a lumped terminal load, which represents a downstream vascular bed and therefore misses the exit over-pressure that jet formation generates at an open exit \citep{krueger2005overpressure,krieg2013modelling,dabiri2009optimal}. The second adopts an open-exit closure from cardiac-valve models, where the unsteady Bernoulli equation sets the pressure drop from the square of the flow rate and an added-mass term in the flow-rate derivative, with coefficients fixed by the valve effective length and cross-sectional area \citep{sun1995transmitral,mynard2012valve}. However, these models have not been validated against directly measured pressure histories across the valve,} \textcolor{black}{and neither closure accounts for the energy that vortex formation carries away at the exit, a loss the present study represents with a vortex-radiation term.} \textcolor{black}{Consequently, the quantitative validity of the unsteady Bernoulli closure for open jet nozzles, particularly those undergoing passive wall deformation, remains unresolved.}


Building on the work of \citet{choi2024mechanism}, this study utilises numerical simulations to quantify wave propagation and reflection, study its influence on the nozzle vortex dynamics, and characterize the energy transfer mechanism (from nozzle elastic potential energy to jet kinetic energy) that leads to thrust enhancement by flexible nozzles. \textcolor{black}{We additionally condense the coupled fluid--structure response into a lumped reduced-order model, closed by an inertial end correction at the nozzle exit, that reproduces the resonance frequency together with the hydrodynamic-impulse and energy histories across the stiffness range, providing a compact and predictive description of the store-and-release dynamics.}

\textcolor{black}{The present study pursues three primary objectives. First, using coupled fluid--structure interaction simulations spanning $Eh = 75$--$500\,\mathrm{N\,m^{-1}}$, we characterise how
stiffness-controlled elastic wave propagation along a passively deforming nozzle governs the pulsed-jet response, from the expansion--contraction timing of the wall to the entrainment, shear-layer roll-up, and vortex ring formation at the exit. Secondly, we quantify the energy exchange between the jet and the compliant wall, tracing the budget from the actuator input through the elastically stored potential energy to the outlet kinetic energy flux. Thirdly, we aim to condense the coupled fluid--structure response into a lumped reduced-order model, closed with physical assumptions, that predicts the resonance frequency and hydrodynamic impulse
directly from the nozzle stiffness and geometry.} These insights have potential applications in enhancing the propulsive performance of underwater vehicles.

\textcolor{black}{The remainder of the paper is organised as follows. Section~2 describes the computational methodology, comprising the coupled fluid--structure interaction framework, the pulsed-jet inlet conditions, and the validation of the numerical setup. Section~3 examines stiffness-controlled wave propagation along the compliant nozzle and its influence on entrainment, shear layer roll-up, and vortex ring formation. Section~4 quantifies the resulting impulse enhancement and the associated energy budget, tracing the exchange between the elastically stored potential energy of the wall and the outlet kinetic-energy flux. Section~5 develops a reduced-order model that condenses the coupled fluid--structure response into a lumped input--output oscillator and assesses its predictions against the simulations. Section~6 summarises the principal conclusions. Supporting derivations, the canonical resonator analogies, the measurement of the store-and-release period, and the all-mode one-dimensional model are collected in appendices~A--F.}


\vspace{-0.4cm}

\section{Computational Methodology}
\label{sec:methodology}
\begin{figure}
	\centering
	\includegraphics[width=1.0\columnwidth]{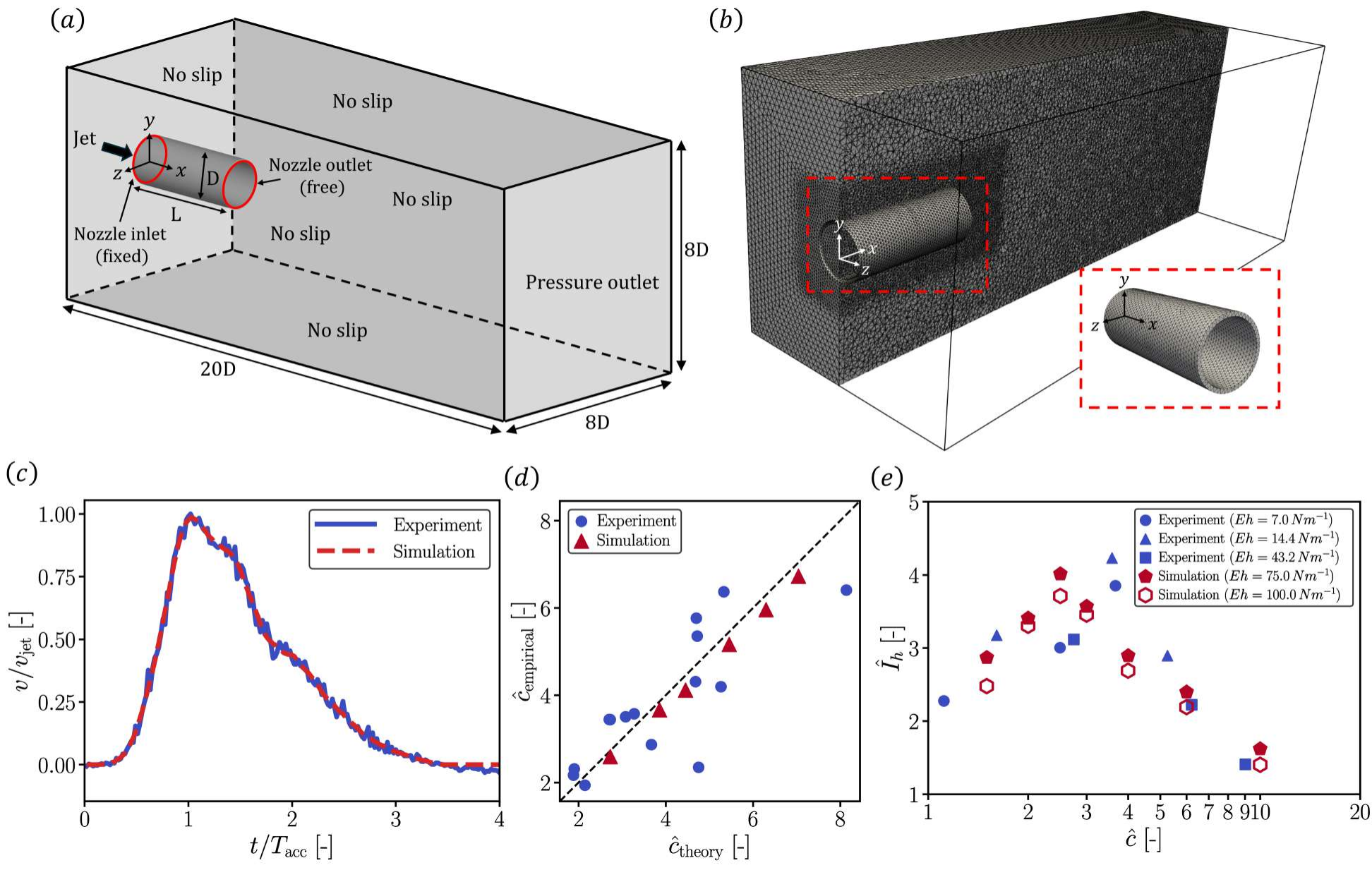}
	\caption{\textit{(a)} Schematic representation of the computational domain with boundary conditions. \textit{(b)} surface and volume mesh for the fluid domain and the inset displaying the mesh considered for the flexible nozzle. \textit{(c)} time history of normalised jet velocity ($v/v_{\mathrm{jet}}$) at the nozzle inlet as a function of normalised time ($t/T_{\mathrm{acc}}$), where $T_{\mathrm{acc}} = 0.05$~s and $v_{\mathrm{jet}} = 0.293~\mathrm{m\,s^{-1}}$. \color{black}{ \textit{(d)} Validation of elastic wave propagation speed along the nozzle wall with experimental and theoretical wave speeds. \textit{(e)} Variation in the normalised hydrodynamic impulse ($\hat{I}_h = I_h/I_{h, \mathrm{ rigid}}$), generated by the flexible nozzle for different dimensionless wave speeds ($\hat{c}$), computed from experiments and simulations.}} 
	\label{V&V}
\end{figure}

The schematic representation of the computational domain along with the associated boundary conditions is presented in figure~\ref{V&V}(a). \textcolor{black}{The cylindrical nozzle of inner diameter $D=15$ mm and length $L=41$ mm (aspect ratio $L/D=2.73$\textcolor{black}{, also $1.3$ and $2.0$ as a geometric test, figure~\ref{fig:resonance}}), corresponding to the experimental configuration of \citet{choi2022flow}, is aligned axially along the $x$-axis, with the base fixed to the flow domain wall and the origin of the Cartesian coordinate system located at the centre of the nozzle inlet.} The fluid domain spans $20D$ along the $x$-axis, $8D$ along the $y$- and $z$-axes. An unstructured grid with tetrahedral elements (shown in figure \ref{V&V}(b)) is used to discretise both the fluid and solid domains to maintain cell quality during mesh deformation around the flexible nozzle. To improve computational efficiency, refinement zones are created where the grid cell size gradually increases with distance from the nozzle. The cell sizes in the solid and neighbouring fluid meshes are kept approximately equal to reduce interpolation error when mapping solutions between the grids. 

A pulsatile jet profile with the peak velocity of $v_{\mathrm{jet}}=0.293~\mathrm{m\,s^{-1}}$ and a jet acceleration time (defined as the time required for the jet velocity to reach $v_{\mathrm{jet}}$) of $T_{\mathrm{acc}}$ = 0.05 s is used for the present study. See figure \ref{V&V}(c). The inlet flow data is obtained from the experimental conditions reported by \citet{choi2024mechanism}. The Reynolds number based on the nozzle diameter and peak jet velocity, defined as $Re=v_{\mathrm{jet}}D/\nu$ (where $\nu$ is the kinematic viscosity of water), is maintained at $Re=4395$. The nozzle is modelled with a uniform wall thickness of $h$ = 1 mm, and solid density $\rho_s=1300~\mathrm{kg\,m^{-3}}$ corresponding to silicone rubber material. The Young's modulus ($E$) is varied from 75 to 500 kPa with a constant Poisson's ratio of 0.4. In the present study, pulsed-jet propulsion through a flexible nozzle with structural stiffness $Eh = 75-500~\mathrm{N\,m^{-1}}$ is analysed and compared with a rigid nozzle ($Eh \to \infty$).

The laminar and incompressible flow is simulated by solving the incompressible Navier-Stokes equations, projected in the arbitrary Lagrangian-Eulerian (ALE) framework, as given by
\begin{subequations}

\begin{align}
\nabla \cdot \mathbf{v}_f &= 0, \label{eq:1a} \\
\frac{\partial \mathbf{v}_f}{\partial t}
+ \big[(\mathbf{v}_f - \mathbf{v}_m)\cdot\nabla\big]\mathbf{v}_f
&= -\frac{1}{\rho}\nabla p + \nu\nabla^2 \mathbf{v}_f. \label{eq:1b}
\end{align}

\end{subequations}
Here, $\rho_f $, $\mu_f$, and $\mathbf{v}_f$ are the density, dynamic viscosity, and the velocity of the fluid, $\mathbf{v}_m$ is the fluid grid point velocity, and $p$ is the fluid pressure. 

The wave propagation dynamics of the elastic nozzle is governed by the structural momentum equation as given by (without any body force)

\begin{align}\label{eq:2}
\rho_s \frac{\partial^2 \mathbf{d}_s}{\partial t^2}-\nabla \cdot \boldsymbol{\sigma}_s=0,
\end{align}

\noindent where $\rho_s$ is the solid density, $\mathbf{d}_s$ is the displacement of the solid, and $\boldsymbol{\sigma}_s$ is the Cauchy stress tensor. For the present study, we have considered a linear-elastic stress-strain relationship:
$\boldsymbol{\sigma}_s = \lambda \, \mathrm{tr}(\boldsymbol{\varepsilon}) \, \mathbf{I} + 2 \mu \, \boldsymbol{\varepsilon}$ and $\boldsymbol{\varepsilon} = \tfrac{1}{2}
\left( \nabla \mathbf{d}_s + (\nabla \mathbf{d}_s)^{T} \right)$, where $\boldsymbol{\varepsilon}$ is the infinitesimal strain tensor, $\mu$ and $\lambda$ represent the shear modulus and the first Lamé parameter, respectively. The operator $\mathrm{tr}(\cdot)$ denotes the trace of a tensor, $\mathbf{I}$ is the second-order identity tensor. It is to be noted that the present linear-elastic results are also compared with those obtained considering a hyper-elastic relationship, and no significant deviation in the coupled structural response is observed. The interface conditions on the fluid-solid interface are given by
\begin{subequations}

\begin{gather}
\mathbf{v}_f = \mathbf{v}_s, \label{eq:3a} \\
\boldsymbol{\sigma}_f \cdot \mathbf{n}_f
+ \boldsymbol{\sigma}_s \cdot \mathbf{n}_s
= \mathbf{0}, \label{eq:3b}
\end{gather}

\end{subequations}
where $\mathbf{n}_f$ and $\mathbf{n}_s$ are the unit outward normal on the interface between fluid and solid.

Two-way coupled fluid–structure interaction simulations are performed using a partitioned strong-coupling approach. In the present open-source fluid–structure interaction framework, the fluid part is simulated using the finite volume method-based {\tt pimpleFoam} solver, available in {\tt OpenFOAM} \citep{weller1998tensorial}, the structural governing equations are solved using finite element method-based code, {\tt CalculiX} \citep{dhondt2017calculix}, and the fluid-solid coupling is established using the library {\tt preCICE} \citep{chourdakis2023openfoam}. For exchanging information at the fluid-solid interface, the radial basis function - thin plate spline (RBF-TPS) interpolation technique is used, and the strong coupling is achieved through the parallel-implicit scheme with the interface quasi-Newton inverse least-squares (IQN-ILS) acceleration method. The spatial and temporal discretisation schemes adopted in this study are second-order accurate.

\textcolor{black}{To quantify the numerical uncertainty of the simulation results and to ensure that they are independent of the spatio-temporal discretisation resolutions, a grid and time-step sensitivity study using the Richardson extrapolation method \citep{celik2008procedure} is carried out, and the results are presented in table \ref{table_mesh_time_step_Richardson_extrapolation}. For the sensitivity study, the maximum nozzle exit velocity measured at a probe located at the center of $Eh = \infty$ and $Eh=150~\mathrm{N\,m^{-1}}$ nozzles is compared using three different grid resolutions (Mesh 1 has $0.5\times10^6$ cells, Mesh 2 has $1.0\times10^6$ cells and Mesh 3 has $2.0\times10^6$ cells), and three time-step resolutions ($\Delta t = 10^{-3}\,\mathrm{s}, 5\times10^{-4}\,\mathrm{s}, 2.5\times10^{-4}\,\mathrm{s}$, using Mesh 2). Thereafter, Mesh 2 and the time step $\Delta t = 5\times10^{-4}$ s (ensuring the maximum Courant number is well below 1) are used for all subsequent simulations, as this combination indicates satisfactory convergence with relative errors $<1.0\%$ for the parameters under consideration.} 

\begin{table}
  \begin{center}
\def~{\hphantom{0}}
  \begin{tabular}{lccccccc}
      \textbf{Parameter} 
      & \multicolumn{3}{c}{$\boldsymbol{(v/v_{\mathrm{jet}})_{\max}}$} 
      & \textbf{Extrapolated}
      & $\boldsymbol{e_{12}}\ \boldsymbol{[\%]}$

      & $\boldsymbol{e_{23}}\ \boldsymbol{[\%]}$

      & $\boldsymbol{e^{\mathrm{extr}}[\%]}$ \\[5pt]

      ~ & \textbf{Mesh 1} &\textbf{ Mesh 2 }& \textbf{Mesh 3} & \textbf{solution} & ~ & ~ & ~ \\[5pt]
      \hline \\

      $Eh=\infty$ 
        & 0.849 & 0.857 & 0.857 
        & 0.857 
        & 0.920 & 0.033 & 0.001 \\

      $Eh=150~\mathrm{N\,m^{-1}}$
        & 1.393 & 1.402 & 1.405 
        & 1.406 
        & 0.680 & 0.205 & 0.089 \\[5pt]

    \hline \\

      ~ & $\boldsymbol{\Delta t = 1\mathrm{e}^{-3}\,\mathrm{s}}$
 & $\boldsymbol{\Delta t = 5\mathrm{e}^{-4}\,\mathrm{s}}$
 & $\boldsymbol{\Delta t = 2.5\mathrm{e}^{-4}\,\mathrm{s}}$
 
      & ~ & ~ & ~ & ~ \\[5pt]
      \hline \\

      $Eh=\infty$
        & 0.851 & 0.857 & 0.858
        & 0.858
        & 0.659 & 0.090 & 0.014 \\

      $Eh=150~\mathrm{N\,m^{-1}}$
        & 1.355 & 1.402 & 1.413
        & 1.416
        & 3.370 & 0.755 & 0.220 \\[5pt]
    \hline
  \end{tabular}
  \vspace{0.3cm}
  \caption{Comparison of peak velocity at nozzle outlet for grid and time-step sensitivity.}
\label{table_mesh_time_step_Richardson_extrapolation}
  \end{center}
\end{table}

In order to validate the present fluid–structure interaction simulation framework, the \textcolor{black}{elastic wave propagation speed along the flexible nozzle wall and the optimal condition for maximizing the impulse generated by the nozzle as a function of the wave-propagation speed are compared with the experimental results of \citet{choi2024mechanism}.} The total hydrodynamic jet impulse $I_h$ is given by $I_h(t)=I_m(t)+I_p(t)$, where $I_m$ and $I_p$ are the momentum impulse and the pressure impulse, respectively \citep{gao2020development}. $I_m(t)=\rho\int_0^t\int_A u_0(r,\tau)\,\left|u_0(r,\tau)\right|\,\mathrm{d}S\,\mathrm{d}\tau$ and $I_p(t)=\int_0^t\int_A \left[p_0(r,\tau)-p_\infty\right]\,\mathrm{d}S\,\mathrm{d}\tau$, where $A$ is the nozzle outlet cross-sectional area, $p_\infty$ is the ambient pressure, and $u_0(r,\tau)$ and $p_0(r, \tau)$ are the axial velocity and pressure at the nozzle exit plane, respectively. \textcolor{black}{The theoretical non-dimensional wave propagation speed ($\hat{c}_{\mathrm{theory}}$) is described by the Moens–Korteweg relation, expressed as $\hat{c}=\sqrt{(E h T_{\mathrm{acc}}^{2})/(\rho_f D L^{2})}$, which depends on the nozzle geometry, stiffness, and inlet jet profile \citep{choi2022flow}. The corresponding experimental and numerical wave speeds ($\hat{c}_{\mathrm{empirical}}$) are obtained from the temporal evolution of the nozzle wall deformation during the initial radial expansion phase, as shown in figure \ref{V&V}(d). The numerical predictions show close agreement with both the analytical scaling and experimental measurements reported by \citet{choi2024mechanism}.}

To compute the impulse at different $\hat{c}$ values, $T_{\mathrm{acc}}$ is varied for \textcolor{black}{ nozzles with stiffness $Eh=75~\mathrm{N\,m^{-1}}$ and $Eh=100~\mathrm{N\,m^{-1}}$}, keeping the nozzle dimensions consistent across all the cases. Additionally, the jet profile is parameterized ($\mathbf{v_f}(t)=\tfrac{v_{\mathrm{jet}}}{2}\!\left[1-\cos\!\left(\tfrac{\pi t}{T_{\mathrm{acc}}}\right)\right]$ when $0\le t<T_{\mathrm{acc}};\;
\tfrac{v_{\mathrm{jet}}}{2}\!\left[1+\cos\!\left(\tfrac{\pi (t-T_{\mathrm{acc}})}{T_{\mathrm{acc}}}\right)\right]$ when $T_{\mathrm{acc}}\le t\le 2T_{\mathrm{acc}}$) to have consistent acceleration and deceleration phases across each case. The computed impulse is normalised using values obtained for a rigid nozzle with the same $T_{\mathrm{acc}}$ value. From figure \ref{V&V}(e), it can be observed that the computed impulse values for different $\hat{c}$ show good agreement with the experimental results and capture the optimal wave speed condition proposed through experiments, with the peak impulse obtained between $\hat{c}$ = 2 to 4. It is worth noting that the impulse values obtained from simulations are not expected to exactly match those from experiments due to differences in nozzle flexibility between the two. 

\vspace{-0.4cm}

\section{\textcolor{black}{Stiffness-controlled wave propagation and vortex formation}}
\label{sec:results}
\begin{figure*}[htbp]
	\centering
	\includegraphics[width=1.0\columnwidth]{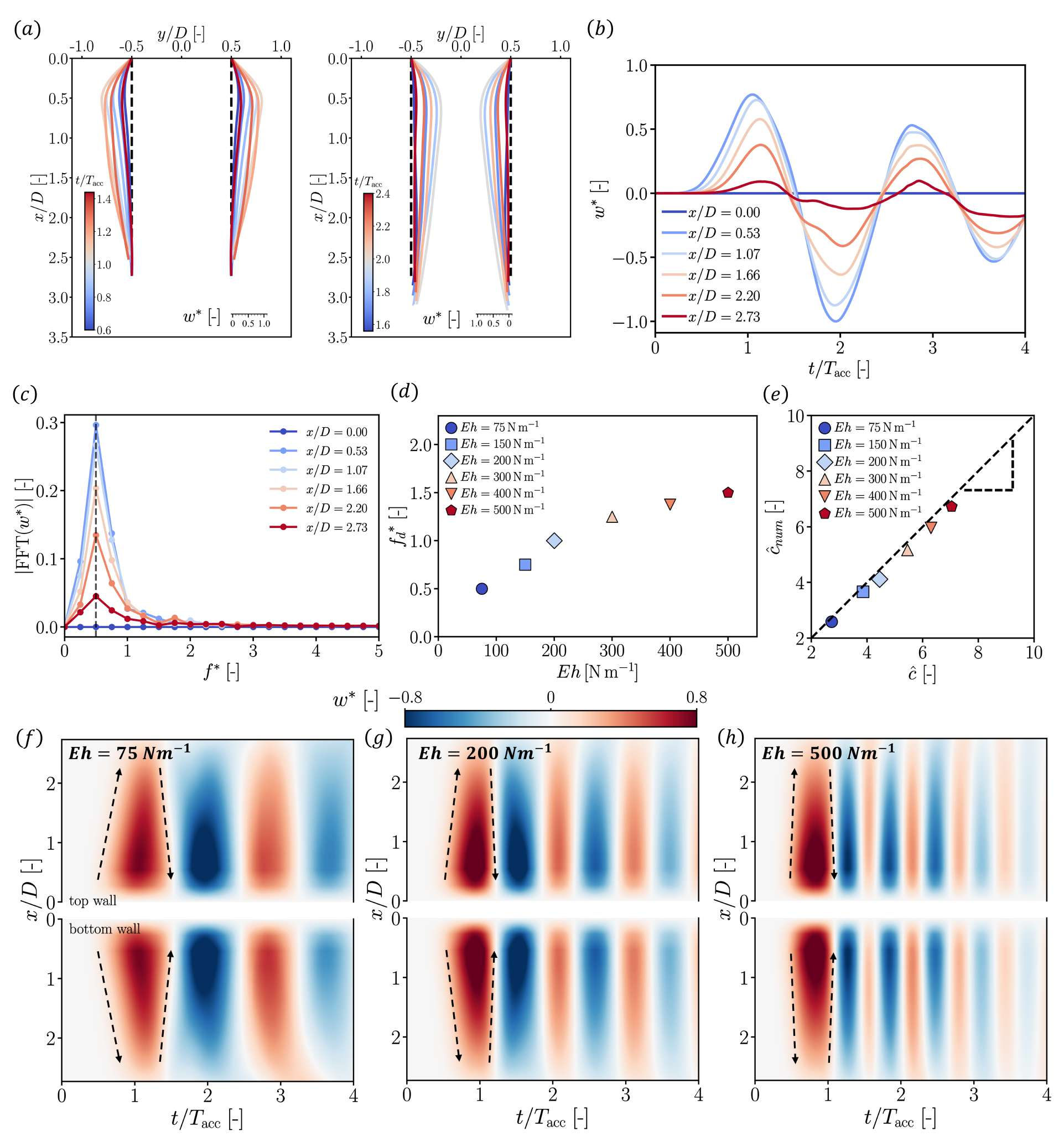}
	\caption{\textit{(a)} Deflection envelope for $Eh=75~\mathrm{N\,m^{-1}}$ nozzle deformation along the $y$ axis during expansion phase (left) and contraction phase (right). \textit{(b)} Temporal evolution of $y$ axis deformations at probes positioned at different axial locations ($x/D$) along the top wall of the $Eh=75~\mathrm{N\,m^{-1}}$ nozzle. \textit{(c)} Frequency spectra for $Eh=75~\mathrm{N\,m^{-1}}$ nozzle obtained from probe deformation measurements. \textit{(d)} Damped natural frequencies $f_d^\star = f_dT_{\mathrm{acc}}$ for different nozzle stiffness. \textit{(e)} Comparison of numerical wave speed with the analytical value for different nozzle stiffness. \textit{(f, g, h)} Spatio-temporal variation of nozzle deformation along $y$ axis ($w^* = w/w_{\mathrm{max}}$) for both top and bottom wall of the nozzle for $Eh=$ $75~\mathrm{N\,m^{-1}}$, $200~\mathrm{N\,m^{-1}}$, and $500~\mathrm{N\,m^{-1}}$ nozzles with $w_{\mathrm{max}} = 0.039D$, $0.014D$, and $0.005D$, respectively.} 
	\label{wave_propagation}
\end{figure*}

To elucidate the underlying mechanisms behind thrust enhancement, pulsed jets through flexible nozzles with stiffness spanning $Eh = 75 - 500~\mathrm{N\,m^{-1}}$ are simulated.
Figure \ref{wave_propagation}(a) illustrates the cross-sectional deflection envelope of the most compliant nozzle ($Eh=75~\mathrm{N\,m^{-1}}$) in the $x$-$y$ plane for the first deformation cycle. The local radial deformations ($w$) are normalised by the peak deflection ($w_{\mathrm{max}}$) during the given interval as $w^* = w/w_{\mathrm{max}}$. As the pulsatile jet initiates and accelerates within the nozzle, it undergoes radial expansion, reaching its peak amplitude before contracting back to its undeformed configuration. During the subsequent jet deceleration phase, the nozzle contracts beyond its undeformed state, and depending on the wave speed, the expansion-contraction cycle continues (refer to supplementary video 1). Since the first deformation cycle aligns with the pulsed jet cycle, it has the largest impact on impulse performance. Consequently, only the first expansion (deformation above the undeformed $y/D$ state) and contraction (deformation below the undeformed $y/D$ state) phases of the nozzle are presented in figure \ref{wave_propagation}(a).

Subsequently, using probes located at different axial positions ($x/D$) along the top wall ($y/D = 0.5$) of the nozzle cross-section in the $x$-$y$ plane, the time histories of $w^*$ 
are presented for $Eh=75~\mathrm{N\,m^{-1}}$. See figure~\ref{wave_propagation}(b). It shows the expansion wave initiating near the inlet probe, then gradually propagating towards the outlet and being reflected, leading to a contraction cycle in the nozzle. The peak deformation during the expansion phase is seen to be in sync with the jet acceleration time $T_{\mathrm{acc}}$.
Next, 
the coupled natural frequency of the most compliant nozzle is calculated using the fast Fourier transform (FFT) of the $w^*$ time histories.
The frequency spectra show a distinct peak at the same non-dimensional frequency $f^* = fT_{\mathrm{acc}}$ across all probe locations. See figure \ref{wave_propagation}(c). 
\textcolor{black}{This dominant frequency $f$ is identified as the damped natural frequency $f_d$, as the nozzle oscillates in a viscous fluid.}
Repeating this analysis for nozzles of varying stiffness yields the damped natural frequencies $f_d^* = f_dT_{\mathrm{acc}}$ (see figure \ref{wave_propagation}(d)), which increase monotonically with stiffness, reflecting the stiffness-controlled dynamics of the flexible nozzle.

To gain a better insight into the wave propagation characteristics, spatio-temporal contours of nozzle deformation (in $x$-$y$ plane) are presented in figures \ref{wave_propagation}(f, g, h). The $y$ axis denotes the axial position along the top and bottom walls of the nozzle, and the contour reflects $w^{*}$ variation. The positive and negative values correspond to expansion (radially outward) and contraction (radially inward), respectively, relative to the undeformed state. Over the complete pulsed jet cycle duration ($4T_{\mathrm{acc}}$), multiple expansion–contraction cycles are observed, with progressively damped amplitudes in subsequent oscillations. In all cases, deformation initiates near the inlet ($x/D = 0$) and propagates downstream toward the outlet ($x/D = 2.73$), where the travelling wave reflects and reverses direction. Consistent with the wave speed scaling $c \propto \sqrt{Eh}$, increasing nozzle stiffness leads to faster wave propagation, causing the expansion wave to complete its cycle in a shorter time. See figures~\ref{wave_propagation}(f, g, h). 
To quantify this behavior, the wave speed associated with the first expansion cycle is extracted from the deflection envelope in figure \ref{wave_propagation}(a), and the resulting numerical non-dimensional wave speeds $\hat{c}_{num}$ are compared against the analytically computed wave speeds 
in figure \ref{wave_propagation}(e). The close agreement confirms that the deformation dynamics are governed by stiffness-controlled wave propagation. Its impact on vortex dynamics is discussed next.

\begin{figure*}[htbp]
	\centering
	\includegraphics[width=1.0\columnwidth]{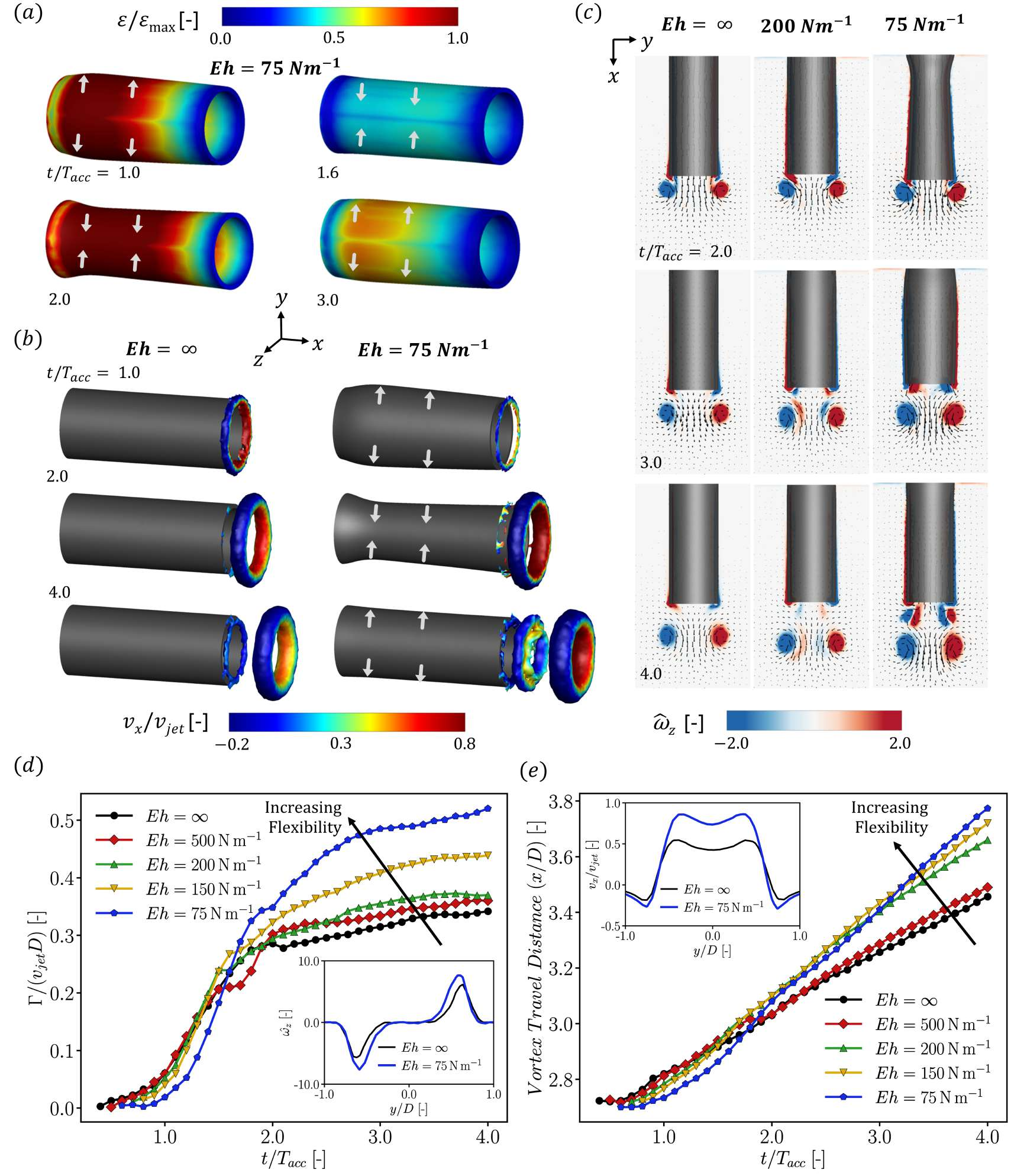}
	\caption{\textit{(a)} Snapshots of 3D nozzle deformation colored by the normalised strain ($\epsilon/\epsilon_{\mathrm{max}}$) for the $Eh=75~\mathrm{N\,m^{-1}}$ nozzle. \textit{(b)} Isosurfaces of the $Q$-criterion ($Q = 200$) snapshots for rigid and flexible nozzles ($Eh=75~\mathrm{N\,m^{-1}}$) colored by the normalised axial component of jet velocity ($v_x$). \textit{(c)} Evolution of jet flow (vorticity and velocity vectors) for different nozzle flexibilities. \textit{(d)} Temporal evolution of primary vortex circulation for different flexibilities and inset presenting the $z$-vorticity at $t/T_{\mathrm{acc}} = 4$ along a line perpendicular to the nozzle axis and passing through the vortex cores. \textit{(e)} Temporal evolution of primary vortex core displacement along nozzle axis for different flexibilities and inset presenting the axial velocity $v_x$ at $t/T_{\mathrm{acc}} = 4$ along a line perpendicular to the nozzle axis and passing through the vortex cores. } 
	\label{vortex_dynamics}
\end{figure*}

The nozzle strain and evolution of the flow structures are investigated as the jet develops and deforms the nozzle. See figure~\ref{vortex_dynamics}. Figure~\ref {vortex_dynamics}(a) shows the normalised strain contours ($\epsilon/\epsilon_{\mathrm{max}}$) of the nozzle, demonstrating the subsequent expansion ($t/T_{\mathrm{acc}} = 1$ and $3$) and contraction ($t/T_{\mathrm{acc}} = 1.6$ and $2.0$) phases of the deformation cycle (see supplementary video 1). Iso-surfaces of the $Q$-criterion  (the second invariant of the velocity gradient tensor), colored by the normalised axial velocity, are shown in figure~\ref{vortex_dynamics}(b) (see supplementary video 2). Due to jet entrainment during the initial expansion phase of the most compliant nozzle ($Eh=75~\mathrm{N\,m^{-1}}$), the roll-up of the shear layer and the formation of the primary vortex ring are delayed, which can be observed at $t/T_{\mathrm{acc}} = 1$. Subsequently, as the nozzle contracts, it accelerates the entrained fluid and amplifies the vortex ring strength and the convection speed, as observed at $t/T_{\mathrm{acc}} = 2$ and $4$. Additionally, the unsteady nozzle deformation promotes the formation of secondary vortex rings (seen at $t/T_{\mathrm{acc}} = 4$), reflecting the repeated acceleration–deceleration imposed on the jet by the travelling deformation wave.

The effect of nozzle flexibility on the vortex dynamics is further investigated through the out-of-plane vorticity ($\omega_z$) contours presented on a $x$-$y$ cross-section plane in figure \ref{vortex_dynamics}(c) (see supplementary video 2). At $t/T_{\mathrm{acc}} = 4$, the primary vortex associated with flexible nozzles convects further downstream and has a higher vortex strength, which is further quantified and presented in figures~\ref{vortex_dynamics}(d, e). Figure~\ref{vortex_dynamics}(d) shows the temporal evolution of primary vortex circulation, defined as $\Gamma = \iint_S \omega_z \, dA$, \textcolor{black}{where $A$ is the area enclosed by the primary vortex. The circulation is computed as the average of the magnitudes of the two counter-rotating primary vortices and is normalised by the peak jet velocity and nozzle diameter.} Consistent with earlier observations, vortex formation is delayed during the nozzle expansion phase and accelerates rapidly during the contraction phase. Furthermore, due to the lower wave speed (as presented in figure \ref{wave_propagation}) for higher flexibility (lower $Eh$) cases, the expansion phase takes longer time to complete and hence in figure \ref{vortex_dynamics}(d), it can be observed that the circulation for $Eh=75~\mathrm{N\,m^{-1}}$ exceeds that of the rigid nozzle case around $t/T_{\mathrm{acc}} = 1.5$ which is the time period for the expansion phase of the nozzle. Once the contraction cycle initiates, the vortex circulation in flexible nozzles increases rapidly above that in rigid nozzles. For the $Eh=75~\mathrm{N\,m^{-1}}$ nozzle, the primary vortex ring circulation at $t/T_{\mathrm{acc}} = 4$ is $52.13\%$ higher compared to rigid nozzle. A similar effect is observed for the primary vortex core displacements along the $x$ axis, as shown in figure \ref{vortex_dynamics}(e), wherein for the $Eh=75~\mathrm{N\,m^{-1}}$ nozzle, the vortex convects $9.0\%$ further downstream compared to the rigid nozzle towards the end of the pulsed jet cycle. The insets in figures \ref{vortex_dynamics}(d,e) further show the $\omega_z$ and axial velocity $v_x$ profiles, sampled across the vortex cores at $t/T_{\mathrm{acc}} = 4$. The insets reveal increases of $24.19\%$ in peak vorticity and $58.02\%$ in peak axial velocity for the flexible nozzle, demonstrating jet amplification due to wave propagation. 
\textcolor{black}{It is noted that the relatively modest increase in vortex travel distance (9\%) compared to the circulation and velocity gains ($>$20\%) is attributed to the delayed vortex formation caused by the prolonged expansion phase. Since the flexible nozzle produces the vortex ring later than the rigid nozzle, the convection distance measured at $t/T_{\mathrm{acc}} = 4$ does not fully reflect the enhanced vortex strength. This difference is expected to grow at later times ($t/T_{\mathrm{acc}} > 4$).}

\vspace{-0.4cm}

\section{\textcolor{black}{Impulse enhancement and the energy budget}}

\begin{figure}
	\centering
	\includegraphics[width=1.0\columnwidth]{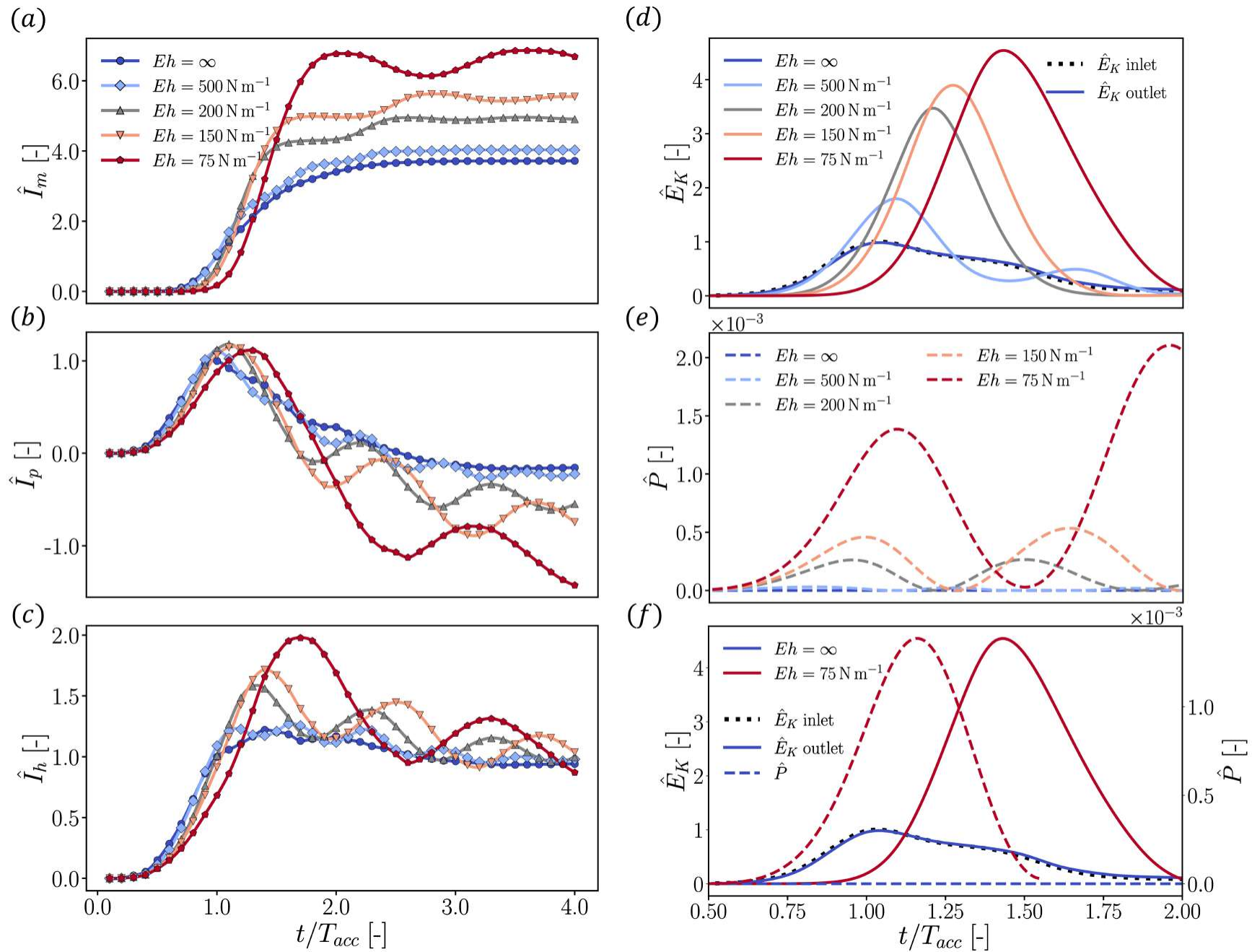}
	\caption{Temporal evolution of \textit{(a)} momentum contribution to impulse \textit{(b)} pressure contribution to impulse and \textit{(c)} total hydrodynamic impulse, normalised by the peak impulse for the rigid nozzle at $t/T_{\mathrm{acc}} = 1$, for different flexibilities. Temporal evolution of \textit{(d)} jet kinetic energy flux at the inlet and outlet, normalised by the peak value at the inlet \textit{(e)} nozzle elastic potential energy, and \textit{(f)} overlay of outlet/inlet kinetic energy flux and elastic potential energy for different flexibilities. }
	\label{impulse_energy}
\end{figure}

The influence of flexibility on impulse generation is quantified in figures~\ref{impulse_energy}(a, b, c). Figure \ref{impulse_energy}(a) shows the momentum contribution to the impulse, $\hat{I}_m$, computed at the nozzle outlet and normalised by the rigid nozzle impulse at $t/T_{\mathrm{acc}} = 1$. The evolution of $\hat{I}_m$ is consistent with the effect of wave propagation on vortex circulation. During the initial expansion phase, the momentum impulse is reduced due to enhanced entrainment \textcolor{black}{(i.e., the incoming jet fluid fills the additional volume created by nozzle expansion)} and the associated decrease in nozzle exit velocity. However, as the nozzle contracts, the entrained fluid is rapidly expelled, resulting in a substantial increase in the outlet momentum. This effect becomes increasingly pronounced with decreasing stiffness, and for the $Eh=75~\mathrm{N\,m^{-1}}$ nozzle, the peak momentum impulse is $84.44\%$ higher compared to the rigid nozzle. Consequently, the nozzle expansion increases the outlet pressure for flexible nozzles, as evident by the higher pressure impulse $\hat{I}_p$ (normalised by the rigid nozzle value at $t/T_{\mathrm{acc}} = 1$). For the $Eh=75~\mathrm{N\,m^{-1}}$ nozzle, the peak pressure impulse is $10.85\%$ higher compared to the rigid nozzle. During contraction, however, the accelerated jet reduces the exit pressure, resulting in a negative contribution to $\hat{I}_p$ that becomes more pronounced with increasing flexibility. Since the magnitude of momentum impulse $\hat{I}_m$ is higher than the pressure impulse  $\hat{I}_p$, the negative effect of nozzle contraction on pressure is offset by the gain in momentum. Thus, the total impulse $\hat{I}_h$ (sum of both momentum and pressure contributions) increases with the nozzle flexibility. See figure \ref{impulse_energy}(c). For the $Eh=75~\mathrm{N\,m^{-1}}$ nozzle, the peak total impulse is $61.92\%$ higher compared to the rigid nozzle.

To examine the energy transfer mechanisms underlying impulse enhancement, the kinetic energy flux at the nozzle inlet and outlet, along with the elastic potential energy stored in the nozzle, are evaluated next. See figures~\ref{impulse_energy}(d, e, f). The kinetic energy flux is defined as $E_{\mathrm{K}}=\int_A \left(\tfrac{1}{2}\rho |\mathbf{v_f}|^2\right)(\mathbf{v_f}\cdot\mathbf{n})\,\mathrm{d}A$, and is normalized by its peak inlet value \textcolor{black}{of the rigid nozzle} to obtain the non-dimensional kinetic energy flux $\hat{E}_K$. See figure \ref{impulse_energy}(d). For flexible nozzles, a clear phase lag is observed between the inlet and outlet kinetic energy, reflecting temporary energy storage during nozzle expansion followed by energetic release during contraction. The elastic potential energy of the nozzle is given by $P=\int_V Z\,\mathrm{d}V$, where $Z = 0.5\sigma_{ij}\epsilon_{ij}$ is the strain energy density, with $\sigma_{ij}$ and $\epsilon_{ij}$ denoting the stress and strain tensors, respectively. The corresponding non-dimensional potential energy, $\hat{P} = 2P/(\pi EhDL)$, is shown in figure \ref{impulse_energy}(e). Two distinct peaks in $\hat{P}$ are observed for flexible nozzles, where the first corresponds to the nozzle expansion phase, while the second arises during contraction. In the energy transfer mechanism, the first peak, corresponding to the expansion phase, is important because it occurs close to $t/T_{\mathrm{acc}}=1$ and governs the subsequent energy release.

To further understand the energy transfer process, both $\hat{E}_k$ and $\hat{P}$ are overlaid in figure \ref{impulse_energy}(f). During the jet acceleration phase ($t/T_{\mathrm{acc}} < 1$), the outlet kinetic energy is lower for the flexible cases than for the rigid case, as the nozzle undergoes expansion and thus temporarily stores a portion of the incoming jet energy as elastic potential energy. Once the inlet kinetic energy reaches its peak and the jet begins to decelerate, the nozzle enters the contraction phase, releasing the stored elastic energy, which is transferred back to the flow and accelerates the jet. As a result, by the end of the expansion–contraction cycle, when the nozzle recovers its undeformed configuration, the outlet kinetic energy is at its peak value and for the $Eh=75~\mathrm{N\,m^{-1}}$ nozzle, this peak kinetic energy is $4.62$ times higher than that of the rigid nozzle, for which the outlet kinetic energy remains comparable to the inlet value throughout the pulse. 
\textcolor{black}{This demonstrates that the flexible nozzle acts as a passive jet amplifier: it extracts additional energy from the actuator, stores it elastically, and releases it to enhance the outlet kinetic energy beyond the rigid counterpart, a mechanism applicable to underwater jet propulsion.}

\vspace{-0.4cm}

\section{Reduced-order modelling}
\label{sec:model}
\color{black}

\begin{figure*}[htbp]
	\centering
	\includegraphics[width=0.72\textwidth]{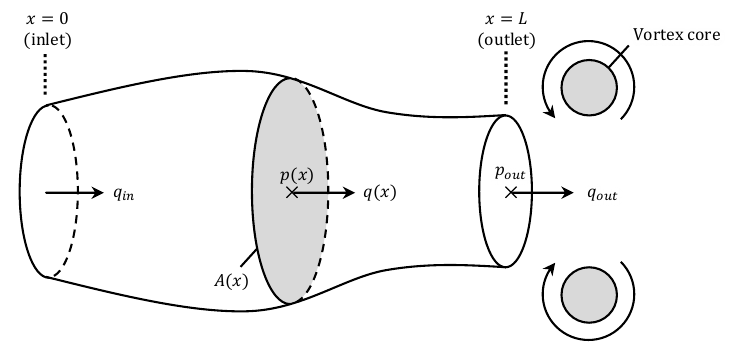}
	\caption{\color{black}Schematic of the one-dimensional compliant-tube model. The pulsatile inflow enters at the inlet ($x=0$) and develops along the tube of cross-sectional area $A(x)$, carrying a volume flow rate $q(x)$ and transmural pressure $p(x)$. At the outlet the efflux leaves with flow rate $q_{\mathrm{out}}$ and over-pressure $p_{\mathrm{out}}$, where the exit shear layer rolls up into the vortex-ring core.}
	\label{fig:schematic}
\end{figure*}

The internal flow of a flexible nozzle can be described by mass conservation and the momentum equation for one-dimensional pulsatile flow in a non-permeable compliant tube (figure~\ref{fig:schematic}) using Reynolds transport theorem~\citep{li2004dynamics,vandevosse2011pulse,hughes1973onedimensional}.

\begin{align}
C\,\frac{\partial p}{\partial t} + \frac{\partial q}{\partial x} &= 0, \\
\frac{\partial q}{\partial t} + \frac{\partial}{\partial x}\!\left(A\,\overline{v_x^{\,2}}\right) + \frac{A}{\rho}\frac{\partial p}{\partial x} &= \oint_{l}\frac{\tau_w}{\rho}\,dl,
\end{align}

\noindent where $p$ is the transmural pressure across the nozzle wall, \textcolor{black}{$p=p_{\mathrm{int}}-p_{\mathrm{ext}}$ taken across the wall, with $p_{\mathrm{ext}}=0$,} $q$ is the volume flow rate, $A$ is the cross-sectional area, $x$ is the axial coordinate, $t$ is time, $\rho$ is the fluid density, $v_x$ is the axial velocity, \textcolor{black}{$\tau_w=\mu\,\partial v_x/\partial r|_{r=R}$ is the wall shear stress, which is negative for a flow retarded by the wall,} $l$ is the perimeter, and $C=\partial A/\partial p$ is the areal compliance of the tube wall.

\begin{figure}[t!]
	\centering
	\includegraphics[width=0.72\columnwidth]{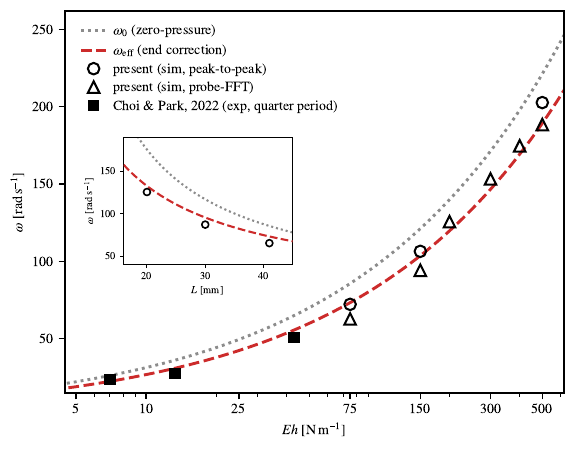}
	\caption{\color{black}Resonance frequency $\omega=2\pi f$ versus structural stiffness $Eh$. The measured wall-ringing frequency from the stored-volume signal (open circles) and the damped natural frequency $\omega_d=2\pi f_d$ extracted from the probe-deflection spectra of figure~\ref{wave_propagation}(\textit{d}) (open triangles) are compared with the zero-pressure prediction $\omega_0$~\eqref{eq:Q_open} and the end-correction prediction $\omega_{\mathrm{eff}}$~\eqref{eq:omega_eff}. \textcolor{black}{The inset repeats the comparison at fixed $Eh=75~\mathrm{N\,m^{-1}}$ against the nozzle length $L=20$, $30$ and $41~\mathrm{mm}$. The closed squares denote the experimental resonance of \citet{choi2022flow}, obtained as $\omega=2\pi/T$ from the quarter-period $T/4$ of their measured exit-velocity histories.}}
	\label{fig:resonance}
\end{figure}

\textcolor{black}{If we scale to $v_x = V\,\check{v}_x$, $p = (\rho V c)\,\check{p}$, $t = (1/\omega)\,\check{t}$, $q = (A_0 V)\,\check{q}$, $A = A_0 \check{A}$, $x = (\lambda/2\pi)\,\check{x}$, $\tau_w = (\eta V/R_0)\,\check{\tau}_w$, and $l = 2\pi R_0\,\check{l}$ \citep{vandevosse2011pulse}, the scaled dimensionless form is}

\begin{align}
\textcolor{black}{\frac{\partial \check{p}}{\partial \check{t}} + \frac{\partial \check{q}}{\partial \check{x}}} &= 0\label{eq:mass_nondim},\\
\textcolor{black}{\frac{\partial \check{q}}{\partial \check{t}} + \frac{V}{c}\frac{\partial}{\partial \check{x}}\left( \check{A} \check{v}_x^{\,2} \right) + \check{A}\frac{\partial \check{p}}{\partial \check{x}}} &= \textcolor{black}{\frac{2}{\alpha^{2}} \oint_{\check{l}} \check{\tau}_w\, d\check{l}}\label{eq:mom_nondim},
\end{align}

\noindent where $V$ is the characteristic velocity, $A_0$ is the reference cross-sectional area, $R_0$ is the reference radius, $\omega$ is the angular frequency, $\eta$ is the dynamic viscosity, $c=\sqrt{E\,h/(\rho\,D)}$ is the Moens--Korteweg wave speed with Young's modulus $E$, wall thickness $h$, and tube diameter $D$, $\alpha^{2}=\rho\,\omega\,R_0^{2}/\eta$ is the squared Womersley number, and $\lambda=2\pi c/\omega$ is the wavelength. \textcolor{black}{Here we neglect the convective and wall-shear terms of the non-dimensional momentum equation, with coefficients $V/c$ and $2/\alpha^{2}$ respectively, on the basis of the following scaling:} for water ($\rho=10^{3}\ \mathrm{kg/m^{3}}$, $\eta=10^{-3}\ \mathrm{Pa\cdot s}$) and the present geometry ($D=0.015\ \mathrm{m}$, $R_0=7.5\times10^{-3}\ \mathrm{m}$), the Moens--Korteweg wave speed $c=\sqrt{Eh/(\rho D)}$ spans $c\approx2.2$--$5.8\ \mathrm{m/s}$ across the stiffness range $Eh=75$--$500~\mathrm{N\,m^{-1}}$, while the characteristic velocity ranges up to the peak jet speed, $V\approx0.2$--$0.3\ \mathrm{m/s}$. The convective term is bounded above by $V/c\lesssim0.3/2.2\approx0.13$ (falling to $\approx0.035$ at the opposite extreme). The wall-shear term is governed by the squared Womersley number $\alpha^{2}=\rho\omega R_0^{2}/\eta$ which attains its smallest value $\alpha^{2}\gtrsim8.8\times10^{2}$, so $2/\alpha^{2}\lesssim2.3\times10^{-3}$. \textcolor{black}{Hence the wall-shear term is negligible, at $O(10^{-3})$, whereas the convective term is only marginal, at $O(10^{-1})$ and reaching $0.13$ at its extreme. The convective term is therefore dropped by choice rather than by scaling, to keep the lumped equation linear so that the effective resonance frequency follows in closed form. Although not pursued in this paper, we find that retaining the convective term produces about a $5\%$ difference in the exit-velocity residual. We also take the area in the coefficients, e.g., the $A$ in $(A/\rho)\,\partial p/\partial x$, as fixed, $A\simeq A_0$, so the error follows the measured strain, which reaches $\Delta A/A_0\approx8\%$ at peak expansion for the most compliant nozzle ($Eh=75~\mathrm{N\,m^{-1}}$, appendix~\ref{app:period}) \textcolor{black}{and falls to $\approx5\%$ and $\approx1\%$ at $Eh=150$ and $500~\mathrm{N\,m^{-1}}$}, holding the error to $O(\Delta A/A_0)\lesssim10\%$, in line with the $\sim4\%$ mean accuracy of the closed model (figure~\ref{fig:resonance}).}



Combining the inviscid, linearised continuity and momentum relations~\eqref{eq:mass_nondim}--\eqref{eq:mom_nondim} and eliminating $p$ gives the undamped wave equation for the flow rate, which in nondimensional form ($\hat{t}=t/T$, $\hat{x}=x/L$, $\hat{q}=q/q_{\max}$\textcolor{black}{, and the matching pressure scale $\hat p=p\,A_0L/(\rho c^{2}Tq_{\max})$}) reads

\begin{equation}
\frac{\partial^{2}\hat{q}}{\partial \hat{t}^{2}}
= \hat{c}^{2}\,\frac{\partial^{2}\hat{q}}{\partial \hat{x}^{2}},
\label{eq:q_wave_nondim}
\end{equation}

\noindent where $\hat{c}\equiv cT/L=T/\tau$ is the nondimensional wave speed ($\tau=L/c$ the wave transit time), $T$ the jet acceleration time, $L$ the nozzle length, and $q_{\max}$ the maximum flow rate. The inlet is driven by the prescribed jet, $q(0,t)=q_{\mathrm{in}}(t)$, and the objective of the reduced-order model is to predict the outlet flow rate $q(1,t)$ at the nozzle exit from the prescribed inlet flow rate $q_{\mathrm{in}}(t)$, from which the exit velocity, the jet impulse and the energy budget follow. Hereafter we drop the hats ($\hat q\to q$, and similarly for $x$ and $t$) and denote the time and axial derivatives by a dot and a prime, respectively.
To prepare an eigenfunction expansion of the forced wave equation~\eqref{eq:q_wave_nondim}, whose inlet is driven by a prescribed, time-dependent flow rate $q_{\mathrm{in}}(t)$, we lift this inlet forcing off the solution so that the remainder $\tilde{q}(x,t)$ obeys a homogeneous inlet condition and can be expanded in the operator's eigenmodes~\citep{herman2015pde}, through the decomposition

\begin{equation}
q(x,t)=q_{\mathrm{in}}(t)+\tilde{q}(x,t),
\label{eq:decomposition}
\end{equation}

\noindent where $q_{\mathrm{in}}(t)$ is spatially uniform and $\tilde{q}(x,t)$ captures the spatial variation, so that $\tilde{q}(0,t)=0$ at the inlet. Substituting the decomposition into Eq.~(\ref{eq:q_wave_nondim}) gives

\begin{equation}
\ddot{\tilde{q}} = \hat{c}^{2}\,\tilde{q}'' - \ddot{q}_{\mathrm{in}},
\label{eq:source}
\end{equation}

\noindent so that the inlet acceleration appears as a source term. The Galerkin method approximates the spatial variation using trial functions $\phi_n$ that satisfy the inlet condition $\phi_n(0)=0$.

\begin{equation}
\tilde{q}(x,t)\approx\sum_n Q_n(t)\phi_n(x).
\label{eq:galerkin}
\end{equation}

\noindent Here $Q_n(t)$ are the time-dependent expansion coefficients and $\phi_n(x)$ the spatial basis functions. We define the latter as the eigenfunctions of the spatial operator $d^2/dx^2$, with the corresponding wavenumbers $\kappa_n$ and natural frequencies $\omega_n$,

\begin{equation}
\phi_n(x)=\sin\!\Big[\frac{(2n-1)\pi x}{2}\Big], \quad \kappa_n=\frac{(2n-1)\pi}{2}, \quad \omega_n=\frac{(2n-1)\pi}{2}\,\hat{c}.
\label{eq:eigenfunctions}
\end{equation}

\noindent These satisfy the requirements of the expansion. At the inlet, $\phi_n(0)=\sin 0=0$. At the outlet, differentiation gives $\phi_n'(x)=\kappa_n\cos(\kappa_n x)$, so that $\phi_n'(1)=\kappa_n\cos[(2n-1)\pi/2]=0$, since the cosine of any odd multiple of $\pi/2$ vanishes, and by mass conservation $\dot p=-q'$~\eqref{eq:mass_nondim} this zero-pressure slope \textcolor{black}{holds the exit pressure fixed in time, $\dot p_{\mathrm{out}}=0$, which with the quiescent initial condition $p_{\mathrm{out}}(0)=0$ is a zero exit pressure $p_{\mathrm{out}}=0$.} As with the inlet, the true, nonzero exit pressure is not imposed on the basis but enters weakly through the retained boundary term, applied when the model is closed below. Differentiating once more gives $\phi_n''(x)=-\kappa_n^2\phi_n(x)$, so the modes decouple the projected equations by orthogonality, whereas a general basis would leave them coupled. Finally, inserting a single free mode $\tilde q=Q_n(t)\phi_n(x)$ into the unforced wave equation $\ddot{\tilde q}=\hat c^2\tilde q''$ gives $\ddot Q_n=-\omega_n^2 Q_n$ with $\omega_n=\kappa_n\hat c$, a frequency-$\omega_n$ oscillator that confirms $\omega_n$ in~\eqref{eq:eigenfunctions} and the odd-harmonic ladder $\omega_n=(2n-1)\omega_1$


Restricting to the first mode (the stiffness scaling $1/\omega_n^2\propto 1/(2n-1)^2$ alone leaves the second mode $\sim1/9$ of the first, and \textcolor{black}{adding the modal forcing coefficient $c_n=4/[(2n-1)\pi]$, which projects the uniform inlet forcing onto mode $n$ (appendix~\ref{app:1dcollapse}), tightens this to $\lvert Q_2\rvert/\lvert Q_1\rvert\propto1/(2n-1)^3\approx1/27$ in the quasi-static limit, while the measured near-resonance value stays $\lesssim0.1$ for the present cases $\hat c\ge2.8$ (figure~\ref{fig:modetrunc})}), we set $\tilde{q}(x,t)\approx Q_1(t)\,\phi(x)$ with $\phi(x)=\sin(\pi x/2)$, the first eigenmode from~\eqref{eq:eigenfunctions}, for which $\phi(0)=0$ and $\phi(1)=1$. Evaluating at the exit gives $Q_1(t)=\tilde{q}(1,t)$ because $\phi(1)=1$, so the modal amplitude is simply the outlet value and we write $\tilde{q}(x,t)\approx \tilde{q}(1,t)\,\phi(x)$. Multiplying by $\phi$ and integrating over $x\in[0,1]$, and using $\int_0^1\phi^2\,\mathrm{d}x=1/2$ for the inertia term and $\int_0^1\phi\,\mathrm{d}x=2/\pi$ for the spatially uniform forcing, the three terms reduce to

\begin{align*}
\int_0^1\!\phi\,\ddot{\tilde{q}}\,\mathrm{d}x &= \tfrac12\,\ddot{\tilde{q}}(1,t),\\
\hat{c}^{2}\!\int_0^1\!\phi\,\tilde{q}''\,\mathrm{d}x &= \hat{c}^{2}\,\underbrace{\tilde{q}'(1,t)}_{\substack{\text{outlet flux gradient}\\ =\,-\dot p_{\mathrm{out}}}}-\tfrac{\pi^2}{8}\,\hat{c}^{2}\,\tilde{q}(1,t),\\
\int_0^1\!\phi\,\ddot{q}_{\mathrm{in}}\,\mathrm{d}x &= \tfrac{2}{\pi}\,\ddot{q}_{\mathrm{in}},
\end{align*}

\noindent where the stiffness term has been integrated by parts once (the three projections are derived term by term in Appendix~\ref{app:projection}). Its inlet end vanishes ($\phi(0)=0$), leaving the axial gradient of the outlet flux $\phi(1)\,\tilde{q}'(1,t)=\tilde{q}'(1,t)$ as the only boundary contribution. By mass conservation $\dot p=-q'$~\eqref{eq:mass_nondim}, this slope is the rate of change of the outlet pressure, $\tilde{q}'(1,t)=-\dot p_{\mathrm{out}}(t)$. Collecting the three and multiplying by two,

\begin{equation}
\ddot{\tilde{q}}(1,t) + \omega_0^2\,\tilde{q}(1,t) = 2\hat{c}^{2}\,\tilde{q}'(1,t) - \frac{4}{\pi}\,\ddot{q}_{\mathrm{in}},
\qquad \omega_0=\frac{\pi}{2}\,\hat{c},
\label{eq:Q_open}
\end{equation}

\noindent the single-mode balance for the outlet flux $\tilde{q}(1,t)$. Its frequency $\omega_0=(\pi/2)\hat{c}$ is that of the zero-pressure condition. Closing~\eqref{eq:Q_open} requires the surviving boundary term, the outlet flux gradient $\tilde{q}'(1,t)$ or equivalently the outlet pressure rate $\dot{p}_{\mathrm{out}}$, in terms of the flow rate. So far, retaining only the first mode has condensed the coupled fluid-structure response into a single input-output oscillator. \textcolor{black}{The higher modes discarded here are computed and their effect examined by the all-mode solution of~\eqref{eq:source} in appendix~\ref{app:1dcollapse}.}

\begin{figure*}[t!]
	\centering
	\includegraphics[width=\textwidth]{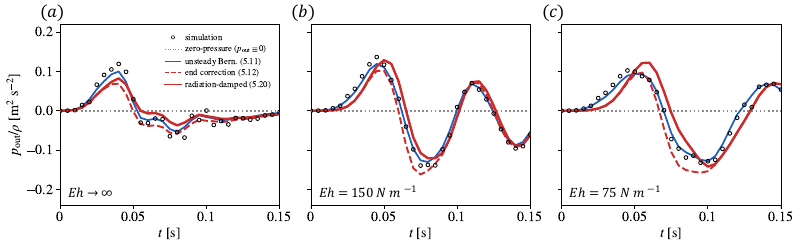}
	\caption{\color{black}Estimation of exit pressure for (\textit{a}) $Eh\to\infty$ (rigid), (\textit{b}) $Eh=150~\mathrm{N\,m^{-1}}$ (less compliant), and (\textit{c}) $Eh=75~\mathrm{N\,m^{-1}}$ (more compliant) nozzles, showing the simulated exit-centreline pressure (circles), the integral form \eqref{eq:bc_integral} (blue), the end-correction closure \eqref{eq:bc_endcorrection} with $L_e=R$ (red dashed), \textcolor{black}{the radiation-damped closure \eqref{eq:closure_loss} (thick red solid),} and the zero-pressure assumption $p_{\mathrm{out}}\equiv0$ (dotted).}
	\label{fig:outlet_pressure}
\end{figure*}

The exit pressure\textcolor{black}{, $p_{\mathrm{out}}(t)=p(x=1,t)$,} varies along with the vortex formation. The measured exit-centreline pressure of the rigid nozzle (figure~\ref{fig:outlet_pressure}(a), circles) increases during jet acceleration owing to the added mass of the external fluid, peaking near $t\approx0.04$~s, and decreases afterward as the jet acceleration falls off and the steady Bernoulli suction takes over. To estimate the exit pressure, the unsteady Bernoulli equation is integrated along the exit-centreline streamline from the quiescent far field (where the velocity and pressure vanish) to the exit plane, yielding~\eqref{eq:bc_integral} (see the detailed derivation in Appendix~\ref{app:bernoulli}). Along the centreline the flow stays essentially irrotational, because the vorticity shed at the nozzle lip is confined to the annular shear layer and vortex core (figure~\ref{fig:schematic}), leaving an irrotational potential core on the axis \citep{krueger2005overpressure}.

\begin{equation}
\underbrace{\frac{p_{\mathrm{out}}(t)}{\rho}}_{\text{exit over-pressure}}=\underbrace{\int_{\mathrm{out}}^{\infty}\frac{\partial u}{\partial t}\,\mathrm{d}x}_{\text{unsteady added mass}}-\underbrace{\tfrac12 u_{\mathrm{out}}^2}_{\text{Bernoulli suction}} ,
\label{eq:bc_integral}
\end{equation}

\noindent where $u_{\mathrm{out}}=q_{\mathrm{out}}/A$ and $x$ is the axial distance measured outward from the exit plane. This integral form (figure~\ref{fig:outlet_pressure}(\textit{a}--\textit{c}), blue solid line) collapses well with the simulations across all three cases, implying that the centreline irrotational assumption remains valid under wall motion.

\textcolor{black}{The centreline distribution of $\partial u/\partial t$ outside the exit plane is self-similar during the early acceleration and follows that of a permeable circular source disk of radius $R$ \citep{rayleigh1896theory,lamb1932hydrodynamics}, so the on-axis flow separates as $u(x,t)=u_{\mathrm{out}}(t)\,\varphi(x)$ with the shape $\varphi$ time-invariant, and that integral collapses onto a single end-correction length $L_e$ multiplying the exit acceleration,}

\begin{equation}
\frac{p_{\mathrm{out}}(t)}{\rho}=L_e\,\frac{\partial u}{\partial t}\bigg|_{\mathrm{out}}-\tfrac12 u_{\mathrm{out}}^2 ,
\qquad L_e=\int_0^\infty\varphi(x)\,\mathrm{d}x ,
\label{eq:bc_endcorrection}
\end{equation}

\noindent an inertial end correction, an effective added length $L_e$ of external fluid accelerating with the exit flow. The source disk gives $\varphi(x)=1-x/\sqrt{x^2+R^2}$ and $L_e=R$ (see detailed derivation in Appendix~\ref{app:selfsimilar}). The zero-pressure outlet of lumped pulsed-jet models is the limiting case $L_e\to0$, for which $p_{\mathrm{out}}\equiv0$. \textcolor{black}{We apply the self-similar form over the whole pulse as a leading-order approximation, examined in figure~\ref{fig:outlet_pressure} and appendix~\ref{app:selfsimilar}.}

The end-correction closure with $L_e=R$ (red dashed in figure~\ref{fig:outlet_pressure}) reasonably follows the simulated exit pressure (circles in figure~\ref{fig:outlet_pressure}). For the rigid nozzle (figure~\ref{fig:outlet_pressure}\textit{a}), \textcolor{black}{the closure and the simulation reach their pressure overshoot at the same time, $t\approx40$~ms, so the timing of the acceleration peak is captured.} The amplitude is underestimated because the real efflux is not purely potential, the shear layer rolling up and entraining external fluid so that the growing vortex adds a centreline acceleration that the attached source-disk column ($L_e=R$) does not carry (appendix~\ref{app:selfsimilar}). For the compliant nozzles (figure~\ref{fig:outlet_pressure}\textit{b},\textit{c}), \textcolor{black}{the closure reproduces the rise-and-fall trend, together with the peak and trough magnitudes. The main discrepancy is now in the timing, the simulated pressure peaking earlier than the closure by about $10$~ms for the most compliant nozzle, where the simulation overshoots at $t\approx45$~ms against $t\approx55$~ms for the $L_e=R$ closure.} The expanding wall makes the exit a non-uniform source, so the on-axis acceleration profile is no longer self-similar and the time-invariant shape assumed in \eqref{eq:bc_endcorrection} no longer holds, so the effective radius and hence $L_e(t)$ drift during the pulse. Though the constant $L_e=R$ estimate has clear theoretical limits, it captures both the trend and the magnitude of the exit pressure reasonably well. We therefore neglect this relatively weak variation of $L_e$ and use the constant rigid value $L_e=R$ as the leading-order end correction.

The exit-pressure model~\eqref{eq:bc_endcorrection} supplies the outlet condition missing from the single-mode balance~\eqref{eq:Q_open}, which needs to be written in terms of the flow rate. Writing $u_{\mathrm{out}}=q/A$ and dropping the quadratic Bernoulli term $u_{\mathrm{out}}^2/2$ \textcolor{black}{as appropriate to the early acceleration phase, where the accelerating exit flow makes the inertial term $L_e\dot u_{\mathrm{out}}$ dominant (its effect is examined by the all-mode model of appendix~\ref{app:1dcollapse}, which retains the unlinearised closure)}, the end-correction pressure becomes $p_{\mathrm{out}}=(\rho L_e/A)\,\dot q|_{\mathrm{out}}$. Differentiating in time and eliminating $\dot p$ with the linearised continuity relation $(A/\rho c^2)\dot p+q'=0$ gives $q'|_{\mathrm{out}}=-(L_e/c^2)\,\ddot q|_{\mathrm{out}}$, and rescaling lengths by $L$ and times by $T$ (so $\hat c=cT/L$) collapses the prefactor $L_e/c^2$ onto $(L_e/L)/\hat c^2$. Since the inlet flow is spatially uniform ($q'=\tilde q'$), \textcolor{black}{while the outlet acceleration inherits both parts of the decomposition~\eqref{eq:decomposition}, $\ddot q|_{\mathrm{out}}=\ddot q_{\mathrm{in}}+\ddot{\tilde q}(1,t)$,} the exit slope becomes

\begin{equation}
\tilde{q}'(1,t)=\textcolor{black}{-\frac{\epsilon}{\hat{c}^{2}}\left[\ddot{\tilde{q}}(1,t)+\ddot{q}_{\mathrm{in}}\right]},\qquad \epsilon\equiv\frac{L_e}{L}=\frac{R}{L},
\label{eq:bc_qform}
\end{equation}

\noindent Substituting the closure~\eqref{eq:bc_qform} into \eqref{eq:Q_open} \textcolor{black}{gives $(1+2\epsilon)\,\ddot{\tilde{q}}(1,t)+\omega_0^{2}\,\tilde{q}(1,t)=-\left(4/\pi+2\epsilon\right)\ddot{q}_{\mathrm{in}}$, and dividing through by $1+2\epsilon$} completes the single-mode model.


\begin{equation}
\ddot{\tilde{q}}(1,t) + \omega_{\mathrm{eff}}^2\,\tilde{q}(1,t) = \textcolor{black}{-\frac{4/\pi+2\epsilon}{1+2\epsilon}}\,\ddot{q}_{\mathrm{in}},
\label{eq:Q_ode}
\end{equation}

\noindent where the shifted resonance frequency \textcolor{black}{follows from the zero-pressure frequency $\omega_0$ already defined in~\eqref{eq:Q_open} as}

\begin{equation}
\omega_{\mathrm{eff}}=\frac{\omega_0}{\sqrt{1+2\epsilon}},
\label{eq:omega_eff}
\end{equation}

\noindent Since $\phi(1)=1$, Eq.~(\ref{eq:decomposition}) gives $q_{\mathrm{out}}=q_{\mathrm{in}}+\tilde{q}(1,t)$. Substituting $\tilde{q}(1,t)=q_{\mathrm{out}}-q_{\mathrm{in}}$ into Eq.~(\ref{eq:Q_ode}) yields

\begin{equation}
\ddot{q}_{\mathrm{out}} + \omega_{\mathrm{eff}}^2 q_{\mathrm{out}} = \omega_{\mathrm{eff}}^2 q_{\mathrm{in}} + \textcolor{black}{\frac{1-4/\pi}{1+2\epsilon}}\,\ddot{q}_{\mathrm{in}}.
\label{eq:qout_exact}
\end{equation}

\noindent The second term on the right-hand side is a small correction, whose magnitude relative to the leading term scales as $|\ddot{q}_{\mathrm{in}}|/(\omega_{\mathrm{eff}}^2|q_{\mathrm{in}}|)\sim 1/\omega_{\mathrm{eff}}^2=(4/\pi^2)(1+2\epsilon)/\hat{c}^{2}$. \textcolor{black}{Multiplied by the feedthrough prefactor $(4/\pi-1)/(1+2\epsilon)\approx0.20$, the factors of $1+2\epsilon$ cancel, so the net contribution is bounded by $(4/\pi-1)(4/\pi^{2})/\hat{c}^{2}\approx0.11/\hat{c}^{2}$ independently of $\epsilon$, the same bound applying to the zero-pressure case $\epsilon=0$. From our experiments and simulations, $\hat{c}\in[1.26,9.99]$, so this contribution ranges from $\sim0.1\%$ to $\sim7\%$ of the leading term. The effect of dropping this term is examined by the all-mode model of appendix~\ref{app:1dcollapse}.} Neglecting it, the lumped input–output ODE is

\begin{equation}
\ddot{q}_{\mathrm{out}} + \omega_{\mathrm{eff}}^2\,\big(q_{\mathrm{out}} - q_{\mathrm{in}}\big) = 0,
\label{eq:lumped_io}
\end{equation}

\noindent which carries both closures. The end-correction model uses $\epsilon=L_e/L$, while setting $\epsilon=0$ recovers the zero-pressure model $\ddot{q}_{\mathrm{out}}+\omega_0^2 q_{\mathrm{out}}=\omega_0^2 q_{\mathrm{in}}$ \textcolor{black}{with the resonance $\omega_0$ of~\eqref{eq:Q_open}}. Both $q_{\mathrm{out}}$ and $q_{\mathrm{in}}$ are functions of time, so $q_{\mathrm{out}}(t)$ follows by solving~\eqref{eq:lumped_io} for a given inlet profile $q_{\mathrm{in}}(t)$.

Figure~\ref{fig:resonance} shows the shifted resonance frequency, $\omega_{\mathrm{eff}}$~\eqref{eq:omega_eff}, compared with the measured ringing frequency directly from the wall stored-volume signal $\Delta V(t)$ (the measurement procedure is detailed in Appendix~\ref{app:period}) and the damped natural frequency $\omega_d=2\pi f_d$ from figure~\ref{wave_propagation}(\textit{d}). The end correction ($\epsilon=R/L$) reproduces the simulated frequency to within $\sim4\%$ on average while the zero-pressure outlet ($p=0$) over-predicts it by $9$--$19\%$ because it omits the boundary added mass. \textcolor{black}{The inset provides an independent geometric test, holding $Eh=75~\mathrm{N\,m^{-1}}$ fixed and varying the nozzle length $L$ over $L/D=1.3$, $2.0$ and $2.7$, so that $\epsilon=R/L$ changes at fixed wave speed. There $\omega_{\mathrm{eff}}$ again follows the simulated resonance while the uncorrected $\omega_0$ over-predicts by $30$--$40\%$, confirming that the boundary added mass acts as a geometric end length $L_e$ rather than a stiffness effect.}

The end correction lowers the natural frequency from $\omega_0$ to $\omega_{\mathrm{eff}}=\omega_0/\sqrt{1+2\epsilon}$, as given by \eqref{eq:omega_eff}. For small $\epsilon$ this reduces to $\omega_{\mathrm{eff}}\approx\omega_0/(1+\epsilon)$, the resonance of a tube of effective length $L(1+\epsilon)=L+L_e$, so the end correction acts simply as an added nozzle length $L_e$ at which the wave reflects, at $x=L+L_e$ rather than at the physical exit $x=L$. This is consistent with figure~\ref{wave_propagation}(\textit{f}--\textit{h}), where the deformation wave reaches the exit, vanishes from the wall for a finite interval ($t/T_{\mathrm{acc}}\lesssim0.5$), and only then re-emerges as a contraction wave propagating back from the exit. This added round trip is $2L_e/c=2\epsilon/\hat c\approx0.1\,T_{\mathrm{acc}}$ for the most compliant nozzle, a small fraction of the pulse that lies within the observed sub-half-cycle ($t/T_{\mathrm{acc}}\lesssim0.5$) gap. The boundary inertia carried by the end correction is therefore what brings the predicted resonance into quantitative agreement with the coupled fluid–structure interaction dynamics across the stiffnesses. \textcolor{black}{The independent soft-nozzle experiments of \citet{choi2022flow} (closed squares in figure~\ref{fig:resonance}), at the same nozzle geometry ($L$ and $D$), corroborate the same $\omega_{\mathrm{eff}}$ trend down to much softer nozzles ($Eh=7$--$43~\mathrm{N\,m^{-1}}$). Their resonance is obtained from a quarter-period measured from the pulse onset ($t=0$) to the instant of maximum nozzle-tip displacement, i.e. when the deformation wave reflects at the tip.}

Equation~\eqref{eq:lumped_io} is a driven harmonic oscillator arising as the single-mode projection of the wave equation for flow rate in a compliant tube. The same mathematical structure, comprising inertia, and a stiffness-restoring term, equally governs an LC transmission line and a spring--mass chain (See detail in Appendix~\ref{app:analogy}). The resonant amplification encoded in~\eqref{eq:lumped_io} is analogous to superpropulsion, whereby elastic energy stored in the compliant wall during the acceleration phase is released in phase with the driving pulse when $\omega_{\mathrm{eff}}$ matches the excitation, yielding an impulse ratio exceeding unity. This store-and-release mechanism has been identified in elastic-projectile throwing \citep{celestini2020contactlayer,giombini2022use}, droplet ejection \citep{prl119108001}, and rapid biological excretion and propulsion \citep{challita2023superpropulsion,choi2026squid}.


\textcolor{black}{We solve~\eqref{eq:lumped_io} as an initial-value problem from rest. Recast as the first-order system $\dot q_{\mathrm{out}}=v$, $\dot v=\omega_{\mathrm{eff}}^2\big(q_{\mathrm{in}}(t)-q_{\mathrm{out}}\big)$, with $q_{\mathrm{in}}(t)$ linearly interpolated from the prescribed inlet waveform of figure~\ref{V&V}(\textit{c}), it is integrated from $q_{\mathrm{out}}=v=0$ by an implicit Runge--Kutta (Radau) scheme (the solver and raw data are available on GitHub, \url{https://github.com/BioFSILab/FlexibleNozzle-PulsedJet-JFM-2026}).}

\begin{figure*}[t!]
	\centering
	\includegraphics[width=\textwidth]{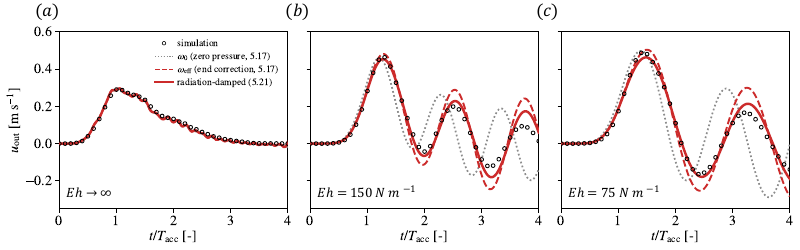}
	\caption{\color{black}Exit-jet velocity $u_{\mathrm{out}}$ history for (\textit{a}) the rigid ($Eh\to\infty$), (\textit{b}) $Eh=150~\mathrm{N\,m^{-1}}$, and (\textit{c}) the most compliant ($Eh=75~\mathrm{N\,m^{-1}}$) nozzles, versus normalised time $t/T_{\mathrm{acc}}$. \textcolor{black}{Circles denote the exit velocity of the numerical simulations, taken as the mass-conserving area average over the exit plane.} The red dashed line is the single-mode model~\eqref{eq:lumped_io} with the geometric end correction ($\omega_{\mathrm{eff}}$, $\epsilon=R/L$), and the grey dotted line the zero-pressure closure without end correction ($\omega_0$, $\epsilon=0$). \textcolor{black}{The thick red solid line adds the radiation damping of~\eqref{eq:lumped_damped}.}}
	\label{fig:velocity_history}
\end{figure*}

\textcolor{black}{Figure~\ref{fig:velocity_history} compares three solutions, the end-corrected resonance $\omega_{\mathrm{eff}}$ ($\epsilon=R/L$), the zero-pressure $\omega_0=\omega_{\mathrm{eff}}\rvert_{\epsilon=0}$ and the radiation-damped model, which will be introduced later, converted to the exit velocity $u_{\mathrm{out}}=q_{\mathrm{out}}/A$, against the flux-averaged outlet-plane velocity of the simulations (the exit pressure of figure~\ref{fig:outlet_pressure} was evaluated on the centreline, whereas all exit quantities from here on are area averages over the outlet plane). For the rigid nozzle every solution collapses onto the simulation, since a rigid wall stores no volume and mass conservation ties the outlet flow to the imposed inlet flow, leaving the exit velocity independent of the closure. Both undamped closures, $\omega_{\mathrm{eff}}$ and $\omega_0$, reproduce the amplitude of the first overshoot to within ${\sim}4\%$. In frequency, however, the two closures diverge, $\omega_{\mathrm{eff}}$ recovering the ringing frequency to within ${\sim}3\%$ whereas the uncorrected $\omega_0$ over-predicts it by ${\sim}11$--$22\%$, which again shows that the zero-pressure closure is missing the boundary added mass. Beyond the first overshoot ($t/T_{\mathrm{acc}}\gtrsim1.5$) the simulated velocity of the compliant nozzles visibly decays while the lumped oscillator rings on undamped. In the open simulation domain each half-cycle ejects a slug and its vortex ring, which convects downstream carrying energy and momentum and does not return. This one-way outflow acts as an effective radiation damping on the exit-plane oscillation (formalised below as the vortex-radiation damping of~\eqref{eq:lumped_damped}), whereas the closed single-degree-of-freedom model traps the ejected energy in the mode. The decay sets in near $t/T_{\mathrm{acc}}\approx1.5$ for the compliant nozzles ($Eh=75$--$500~\mathrm{N\,m^{-1}}$), coinciding with vortex-ring detachment, and strengthens as $Eh$ decreases. Successive local maxima of the exit velocity fall by a factor $0.43$ for $Eh=150~\mathrm{N\,m^{-1}}$ and $0.34$ for $Eh=75~\mathrm{N\,m^{-1}}$, while the rigid nozzle decays monotonically with no ringing to damp.}

\textcolor{black}{The closure~\eqref{eq:bc_endcorrection} is a reversible potential response, so the oscillator~\eqref{eq:lumped_io} conserves the ejected energy and rings on undamped. Multiplying~\eqref{eq:lumped_io} (in $u=q/A$) by $\dot u_{\mathrm{out}}$ and grouping the total derivatives gives}

\textcolor{black}{\begin{equation}
\frac{\mathrm d}{\mathrm dt}\big(P+E_{K}\big)=\omega_{\mathrm{eff}}^{2}\,u_{\mathrm{in}}\,\dot u_{\mathrm{out}},
\qquad P=\tfrac12\dot u_{\mathrm{out}}^{2},\qquad E_{K}=\tfrac12\omega_{\mathrm{eff}}^{2}u_{\mathrm{out}}^{2},
\label{eq:energy_balance}
\end{equation}}
\textcolor{black}{\noindent the two halves of the conserved modal energy $E=P+E_{K}$, namely $E_{K}\propto u_{\mathrm{out}}^{2}$ is the kinetic energy of the oscillating exit column, and $P=k\,x^{2}/2$ the elastic energy stored in the compliant wall, with dilation $x=\Delta A$ and stiffness $k=1/C=Eh/(A_{0}D)$ the inverse of the areal compliance $C=\partial A/\partial p$, built from the membrane stiffness $Eh$, the reference cross-sectional area $A_{0}$ and the tube diameter $D$, so $Eh$ sets both $C\propto1/Eh$ and $\omega_{\mathrm{eff}}\propto\sqrt{Eh}$. The inlet drive $\omega_{\mathrm{eff}}^{2}u_{\mathrm{in}}\dot u_{\mathrm{out}}$ pumps energy in but no term removes it, so the mode rings on. The simulations instead decay beyond the first overshoot ($t/T_{\mathrm{acc}}\gtrsim1.5$, figure~\ref{fig:velocity_history}) because the open domain carries energy away, as each ejection half-cycle rolls the efflux into a vortex ring that convects downstream and never returns, while the suction half-cycle re-ingests ambient fluid as an attached, irrotational potential sink that stores and returns its energy reversibly.}

\textcolor{black}{We therefore represent this energy sink as an exit over-pressure whose magnitude is fixed by the momentum theorem for the discharged jet. The reversible closure~\eqref{eq:bc_endcorrection} is corrected at the exit by two dissipative pressure terms, one for each flow direction, a radiation over-pressure $\Delta p_{\mathrm{rad}}$ during ejection ($u_{\mathrm{out}}>0$) and an entrance loss $\Delta p_{\mathrm{suc}}$ during suction ($u_{\mathrm{out}}<0$). Adding both corrections to~\eqref{eq:bc_endcorrection},}

\textcolor{black}{\begin{equation}
\begin{gathered}
\frac{p_{\mathrm{out}}(t)}{\rho}=L_{e}\,\dot u_{\mathrm{out}}-\tfrac12 u_{\mathrm{out}}^{2}+\frac{\Delta p_{\mathrm{rad}}}{\rho}-\frac{\Delta p_{\mathrm{suc}}}{\rho},\\[2pt]
\frac{\Delta p_{\mathrm{rad}}}{\rho}=\tfrac12\,K_{\mathrm{out}}\,u_{\mathrm{out}}^{2}\,H(u_{\mathrm{out}}),\qquad
\frac{\Delta p_{\mathrm{suc}}}{\rho}=\tfrac12\,K_{\mathrm{in}}\,u_{\mathrm{out}}^{2}\,H(-u_{\mathrm{out}}),
\end{gathered}
\label{eq:closure_full}
\end{equation}}
\textcolor{black}{\noindent with $H$ the Heaviside step and $K_{\mathrm{out}}$, $K_{\mathrm{in}}$ the loss coefficients scaling the ejection radiation over-pressure $\Delta p_{\mathrm{rad}}$ and the suction over-pressure $\Delta p_{\mathrm{suc}}$, respectively. During ejection the efflux separates at the sharp lip and issues as a free jet into effectively unbounded quiescent fluid. The radiation over-pressure $\Delta p_{\mathrm{rad}}$ cancels the Bernoulli suction of~\eqref{eq:bc_endcorrection} while the jet issues. By the momentum theorem for such a discharge (the Borda--Carnot limit of a sudden expansion), the emerging jet sits at ambient pressure, so its velocity head is carried away with the vortex ring rather than recovered, giving $K_{\mathrm{out}}=1$ \citep{borda1766memoire,carnot1783essai}. During suction $\Delta p_{\mathrm{suc}}$ is the entrance loss of a re-entrant intake, where the inflow converges on the lip as an attached potential sink that recovers its head apart from that loss, with $K_{\mathrm{in}}\approx0.78$ the entrance-loss coefficient \citep{idelchik2007handbook,crane2013flow}. This intake term acts only on the weak suction ($u_{\mathrm{ring}}/u_{\mathrm{pk}}\lesssim0.4$), so it is of order $K_{\mathrm{in}}(u_{\mathrm{ring}}/u_{\mathrm{pk}})^{2}\approx0.1$ of the ejection-phase head and shifts the exit-velocity history by less than $0.5\%$. Setting $K_{\mathrm{out}}=1$ (full radiation) and $K_{\mathrm{in}}=0$ hereafter, the completed exit closure is}

\textcolor{black}{\begin{equation}
\frac{p_{\mathrm{out}}(t)}{\rho}=L_{e}\,\dot u_{\mathrm{out}}-\tfrac12 u_{\mathrm{out}}^{2}\,H(-u_{\mathrm{out}}),
\label{eq:closure_loss}
\end{equation}}
\textcolor{black}{\noindent so the Bernoulli suction survives only on inflow ($u_{\mathrm{out}}<0$ at the exit), where the head is recovered (stored as wall elastic energy), and is spent on ejection ($u_{\mathrm{out}}>0$), where it radiates with the ring. Evaluated on the measured exit velocity, this radiation-damped closure (thick red line in figure~\ref{fig:outlet_pressure}) follows the simulated exit pressure and corrects the trough depths that the reversible closure~\eqref{eq:bc_endcorrection} overshoots for the compliant nozzles, which supports cancelling the ejection suction.}

\textcolor{black}{The single-mode balance~\eqref{eq:Q_open} couples to the flow only through the outlet-pressure rate, $\tilde q'(1,t)=-\dot p_{\mathrm{out}}$. Of the four pressure terms in the completed closure~\eqref{eq:closure_full}, the inertial $L_e\dot u_{\mathrm{out}}$ already underlies the undamped oscillator~\eqref{eq:lumped_io}, the reversible Bernoulli head $-u_{\mathrm{out}}^{2}/2$, a head the attached exterior flow stores and returns rather than loses, stays dropped as in the flow-rate form~\eqref{eq:bc_qform} (if reinstated, it would cancel $\Delta p_{\mathrm{rad}}$ on ejection and the oscillator would conserve energy that leaves with the ring), and the suction over-pressure is zero with $K_{\mathrm{in}}=0$, as discussed above. \textcolor{black}{The reason the two heads must not be combined is that they belong to different balances. Dropping the convective term from~\eqref{eq:mom_nondim} leaves the linearised tube with a single exit energy channel, $\mathrm{d}E/\mathrm{d}t=p_{\mathrm{in}}q_{\mathrm{in}}-p_{\mathrm{out}}q_{\mathrm{out}}$, and no kinetic-energy flux term of its own. The jet nonetheless carries $\tfrac12\rho A u_{\mathrm{out}}^{3}$ away with each slug, so that flux can only leave the resonator through $p_{\mathrm{out}}q_{\mathrm{out}}$. The quantity entering the projection is therefore an effective, energy-transporting exit pressure, which carries the velocity head outward, and not the static exit pressure of~\eqref{eq:closure_loss}, which the reversible head cancels. Figure~\ref{fig:outlet_pressure} tests the latter against the simulated exit pressure, while the oscillator is built on the former.} What remains, therefore, is the radiation over-pressure $\Delta p_{\mathrm{rad}}/\rho=u_{\mathrm{out}}^{2}H(u_{\mathrm{out}})/2$, kept for its character rather than its size, as the one irreversible term it is the only sink the energy balance~\eqref{eq:energy_balance} can acquire (\eqref{eq:energy_damped} below). Differentiating the retained term, $\mathrm{d}\!\big(u^{2}H(u)/2\big)/\mathrm{d}t=u\dot u\,H(u)$ (differentiating the step itself yields a term carrying the factor $u_{\mathrm{out}}^{2}$, which vanishes at the crossing $u_{\mathrm{out}}=0$ where the step switches), a term proportional to $\dot u_{\mathrm{out}}$, a damping. Carrying it through the identical continuity-plus-Galerkin reduction that took~\eqref{eq:bc_endcorrection} to~\eqref{eq:bc_qform} and then to~\eqref{eq:Q_ode} reproduces the inertial map with the same prefactor $2/[(1+2\epsilon)L]$ (the steps are detailed in appendix~\ref{app:projection}), so~\eqref{eq:lumped_io} acquires an ejection-gated nonlinear damping, which written in the flow rate ($q_{\mathrm{out}}=A\,u_{\mathrm{out}}$) reads}

\textcolor{black}{\begin{equation}
\ddot q_{\mathrm{out}}+\frac{2}{(1+2\epsilon)LA}\,q_{\mathrm{out}}H(q_{\mathrm{out}})\,\dot q_{\mathrm{out}}+\omega_{\mathrm{eff}}^{2}\big(q_{\mathrm{out}}-q_{\mathrm{in}}\big)=0.
\label{eq:lumped_damped}
\end{equation}}
\textcolor{black}{\noindent Rewritten in the exit velocity $u=q/A$ and multiplied by $\dot u_{\mathrm{out}}$, \eqref{eq:lumped_damped} adds the sink to the energy balance~\eqref{eq:energy_balance}, where it was absent,}

\textcolor{black}{\begin{equation}
\frac{\mathrm d E}{\mathrm dt}=\omega_{\mathrm{eff}}^{2}u_{\mathrm{in}}\dot u_{\mathrm{out}}-\frac{2}{(1+2\epsilon)L}\,u_{\mathrm{out}}H(u_{\mathrm{out}})\,\dot u_{\mathrm{out}}^{2},
\label{eq:energy_damped}
\end{equation}}
\textcolor{black}{\noindent whose second term is non-negative ($\ge0$) and active only on ejection. It is the radiated vortex kinetic energy, carried from the resonator by the ejected velocity head. Read against the canonical damped oscillator $\ddot q+2\zeta\omega\,\dot q+\omega^{2}q=0$, where the damping ratio $\zeta$ is the fraction of critical damping and sets the decay per ringing cycle, the damping term of~\eqref{eq:lumped_damped} plays the role of $2\zeta\omega_{\mathrm{eff}}\,\dot q_{\mathrm{out}}$ with $2\zeta\omega_{\mathrm{eff}}=2u_{\mathrm{out}}H(u_{\mathrm{out}})/[(1+2\epsilon)L]$. The ratio is therefore $\zeta=u_{\mathrm{out}}H(u_{\mathrm{out}})/[(1+2\epsilon)L\,\omega_{\mathrm{eff}}]$, and since $\omega_{\mathrm{eff}}\propto c/L$ the nozzle length cancels, leaving $\zeta\propto u_{\mathrm{out}}/c$. This ratio vanishes for a rigid nozzle, where $c\to\infty$ and there is no ringing to damp, and grows with compliance (decreasing $c$), consistent with the stiffness-dependent decay of figure~\ref{fig:velocity_history}.}

\textcolor{black}{We integrate~\eqref{eq:lumped_damped} exactly as~\eqref{eq:lumped_io}, from rest by the same Radau scheme, with the gated damping $2q_{\mathrm{out}}H(q_{\mathrm{out}})\dot q_{\mathrm{out}}/[(1+2\epsilon)LA]$ carried in the $\dot v$ equation. Because that gate is non-smooth at $q_{\mathrm{out}}=0$, we checked it against a variable-order backward-differentiation-formula (BDF) solve and an event-detected solve that halts at each zero crossing, both of which reproduce $u_{\mathrm{out}}$ to within $10^{-4}\%$ for every nozzle.}

\textcolor{black}{The radiation-damped model~\eqref{eq:lumped_damped} is assessed on the exit-velocity history against the simulations, one panel per stiffness in figure~\ref{fig:velocity_history}(\textit{a}--\textit{c}). The zero-pressure closure ($\omega_0$, grey dotted line) reproduces the overall exit-velocity magnitude and is accurate for the rigid nozzle, which does not ring. At $Eh=75~\mathrm{N\,m^{-1}}$ (figure~\ref{fig:velocity_history}(\textit{c})) all three closures capture the first overshoot to within $5\%$, and the second overshoot separates them. The zero-pressure closure over-predicts the ringing frequency and falls out of phase, its second exit-velocity overshoot arriving at $t/T_{\mathrm{acc}}\approx2.9$ against $\approx3.2$ in the simulation, about $0.3\,T_{\mathrm{acc}}$ ($\approx16$~ms) early. The end correction $\omega_{\mathrm{eff}}$ (red dashed line) restores this timing, bringing the second overshoot to the simulated $t/T_{\mathrm{acc}}\approx3.2$ (figure~\ref{fig:velocity_history}(\textit{c})), but by conserving the ejected energy instead of radiating it with the vortex ring it leaves that overshoot undamped, its peak velocity $81\%$ above the simulated second overshoot ($45\%$ at $Eh=150~\mathrm{N\,m^{-1}}$) and barely decayed from the first, a peak ratio $\approx0.6$ against the simulated $\approx0.34$. The radiation-damped model~\eqref{eq:lumped_damped} (thick red solid line) reduces this artefact (the timing is right and, as in the simulation, the second overshoot is damped), at the cost of a first overshoot low by ${\sim}5\%$.}

\begin{figure*}[t!]
	\centering
	\includegraphics[width=\textwidth]{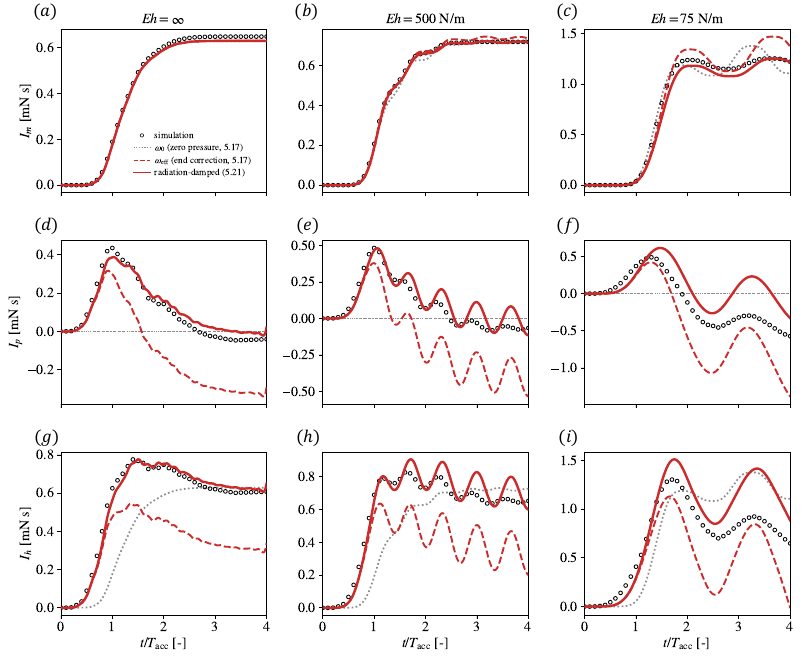}
	\caption{\color{black}(\textit{a}--\textit{c}) Momentum impulse $I_m$, (\textit{d}--\textit{f}) pressure impulse $I_p$ and (\textit{g}--\textit{i}) total hydrodynamic impulse $I_h=I_m+I_p$ versus normalised time, one column per stiffness ($Eh\to\infty$, $500$ and $75~\mathrm{N\,m^{-1}}$, left to right), in dimensional units so that no normalisation can mask an amplitude error. Circles denote the numerical simulations, with the exit velocity and pressure taken as area averages over the exit plane. The red dashed line is the single-mode model~\eqref{eq:lumped_io} closed by the reversible end correction~\eqref{eq:bc_endcorrection}, the grey dotted line the zero-pressure outlet ($p_{\mathrm{out}}\equiv0$, so $I_p\equiv0$), and the thick red solid line the radiation-damped model~\eqref{eq:lumped_damped} closed by~\eqref{eq:closure_loss}. Every closure uses the geometric $L_e=R$.}
	\label{sif_impulse}
\end{figure*}

\textcolor{black}{Figure~\ref{sif_impulse} shows (\textit{a}--\textit{c}) the momentum impulse $I_m$, (\textit{d}--\textit{f}) the pressure impulse $I_p$ and (\textit{g}--\textit{i}) the total $I_h=I_m+I_p$ against the simulations, one column per stiffness, in dimensional units, so that the amplitudes are compared directly. As in the velocity of figure~\ref{fig:velocity_history}, every closure (the zero-pressure $\omega_0$ and end-corrected $\omega_{\mathrm{eff}}$ forms of~\eqref{eq:lumped_io} and the radiation-damped~\eqref{eq:lumped_damped}) captures the rigid-nozzle $I_m$, since the impulse is the time integral of the exit velocity. At $Eh=75~\mathrm{N\,m^{-1}}$ the zero-pressure $\omega_0$ falls out of phase as it did in the velocity and slightly under-predicts the first overshoot ($t/T_{\mathrm{acc}}\approx1.9$ against the simulated $2.1$), yet its second overshoot ($t/T_{\mathrm{acc}}\approx3.3$ against $3.6$) comes out larger than measured ($1.38$ against $1.25~\mathrm{mN\,s}$), the energy it fails to radiate returning a cycle later. The end correction $\omega_{\mathrm{eff}}$ restores the phase but, conserving that same energy, over-predicts the level, and the radiation damping removes the excess without degrading the ringing.}

\textcolor{black}{The closures differ most in the pressure impulse $I_p$ of figure~\ref{sif_impulse}(\textit{d}--\textit{f}). For the rigid nozzle the end-corrected but reversible closure~\eqref{eq:bc_endcorrection} tracks $I_p$ up to the overshoot and falls below it beyond $t/T_{\mathrm{acc}}\approx1$, drifting negative to average $-0.32~\mathrm{mN\,s}$ over $t/T_{\mathrm{acc}}=3$--$4$ against $-0.04~\mathrm{mN\,s}$ in the simulation. Omitting the radiation over-pressure leaves its Bernoulli suction acting on both strokes, a drift the radiation-damped closure~\eqref{eq:closure_loss} removes, returning $I_p$ to zero to within $0.01~\mathrm{mN\,s}$ over the same window and recovering its peak to within $0.05~\mathrm{mN\,s}$. The radiation-damped closure holds for the compliant nozzles as well, reproducing the positive exit pressure that follows the overshoot over $t/T_{\mathrm{acc}}\approx1$--$1.3$, but over-predicting the swings of $I_p$ about its mean drift. At $Eh=500~\mathrm{N\,m^{-1}}$ (figure~\ref{sif_impulse}(\textit{e})) the model rings between successive extrema by up to $0.26~\mathrm{mN\,s}$ over $t/T_{\mathrm{acc}}\approx1.4$--$3.0$, against at most $0.16~\mathrm{mN\,s}$ in the simulation, and at $Eh=75~\mathrm{N\,m^{-1}}$ (figure~\ref{sif_impulse}(\textit{f})) the rebound after the trough climbs to $+0.23~\mathrm{mN\,s}$ at $t/T_{\mathrm{acc}}\approx3.3$ where the simulation recovers only to $-0.30$. The net drift itself still tracks the simulation. For the most compliant nozzle ($Eh=75~\mathrm{N\,m^{-1}}$) the initial pressure rise up to the overshoot ($t/T_{\mathrm{acc}}\lesssim1.3$) departs from the measurement (the model lags the rise and runs low by up to $0.13~\mathrm{mN\,s}$ at $t/T_{\mathrm{acc}}\approx0.9$, $0.17$ against $0.30~\mathrm{mN\,s}$ in the simulation), because the flexible wall violates the self-similar assumption behind the constant end correction $L_e$~\eqref{eq:bc_endcorrection}. The compliant nozzle does not merely issue flow at the exit but expands along its wall, so the source is no longer confined to the exit plane and the uniform source-disk picture (see figure~\ref{sif_source_disk} in appendix~\ref{app:selfsimilar}) no longer holds. Letting $L_e$ vary in time (large during the early expansion and decreasing towards $R$ thereafter) would improve the estimate, which we leave outside the present scope. The constant-$L_e$ closure therefore over-predicts the pressure-impulse overshoot of figure~\ref{sif_impulse}(\textit{f}) in both its timing ($\approx0.2\,T_{\mathrm{acc}}$, ${\approx}9$~ms late) and its peak ($+26\%$). This error propagates to the total impulse (figure~\ref{sif_impulse}(\textit{g}--\textit{i})). At the overshoot of the most compliant nozzle ($Eh=75~\mathrm{N\,m^{-1}}$) the pressure impulse is $\approx18\%$ of the momentum impulse, but it is what creates the overshoot, the zero-pressure closure $\omega_0$ of~\eqref{eq:lumped_io} (grey dotted line) carrying $p_{\mathrm{out}}\equiv0$ and showing no early overshoot for the rigid and $Eh=500~\mathrm{N\,m^{-1}}$ nozzles, its total impulse peaking only with $I_m$. The over-predicted $I_p$ swings therefore lift the $I_h$ overshoot at $t/T_{\mathrm{acc}}\approx1.7$ by $16\%$ ($1.51$ against the simulated $1.30~\mathrm{mN\,s}$), while $I_m$ averaged over $t/T_{\mathrm{acc}}=3$--$4$ stays within $2.3\%$ of the simulation. Because the over-predicted pressure component both raises and delays the total-impulse overshoot, a quantitative prediction of it hinges on refining the pressure closure, in particular the compliance-dependent, time-varying $L_e$ noted above.}

\begin{figure*}[t!]
	\centering
	\includegraphics[width=\textwidth]{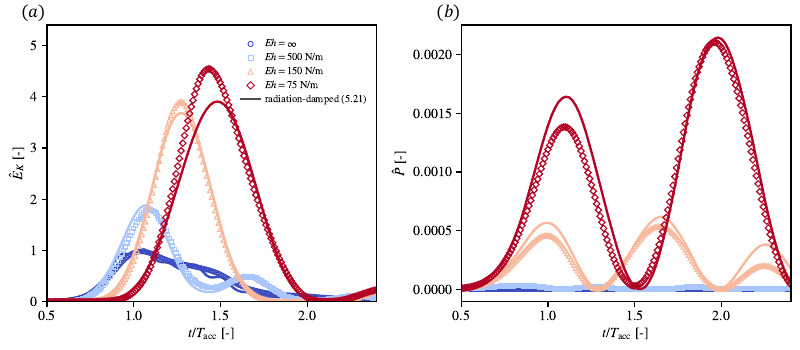}
	\caption{\color{black}Energy budget for the rigid and flexible nozzles. \textit{(a)} Outlet kinetic-energy flux \textcolor{black}{$\hat E_K\propto q_{\mathrm{out}}^3$} (normalised by the peak inlet value) and \textit{(b)} nozzle elastic potential energy $\hat P$, versus normalised time $t/T_{\mathrm{acc}}$. Markers denote the numerical simulations. \textcolor{black}{Solid lines the radiation-damped model~\eqref{eq:lumped_damped}, with the end correction $L_e=R$}.}
	\label{sif_energy}
\end{figure*}

The corresponding energy budget reproduces the same store-and-release signature (figure~\ref{sif_energy}). We track the outlet kinetic-energy flux \textcolor{black}{$\hat E_K\propto q_{\mathrm{out}}^3$} and the nozzle elastic potential energy \textcolor{black}{$\hat P=2P/(\pi EhDL)$, the membrane strain energy $P=\int_V Z\,\mathrm{d}V$ with strain-energy density $Z=\tfrac12\sigma_{ij}\epsilon_{ij}$, evaluated from the solid solver (\S\ref{sec:results})}. Both strengthen by orders of magnitude from the rigid to the most compliant nozzle\textcolor{black}{, as the stiffness scalings anticipate. At a given driving pressure the tube law sets the wall dilation as $\Delta A\propto1/(Eh)$ and the strain energy scales as $P\propto Eh\,\Delta A^{2}$, so the normalised store obeys $\hat P\propto\Delta A^{2}\propto(Eh)^{-2}$ and vanishes in the rigid limit. The kinetic-energy flux is cubic in the flow rate, so its factor of $4.6$ between the most compliant and the rigid nozzle corresponds to a factor of only ${\approx}1.7$ in peak exit velocity.} \textcolor{black}{In the simulations (markers in figure~\ref{sif_energy}), }the kinetic-energy flux peaks with a phase lag that lengthens with compliance, while the elastic energy follows the characteristic two-peak history of expansion followed by contraction. \textcolor{black}{The radiation-damped model (solid lines) captures these trends and matches the first kinetic-energy peak to within $4.0\%$ for the rigid nozzle, $4.5\%$ for the flexible $Eh=500~\mathrm{N\,m^{-1}}$ nozzle and $5.6\%$ for the flexible $Eh=150~\mathrm{N\,m^{-1}}$ nozzle, and under-predicts it by $14\%$ for the most compliant nozzle ($Eh=75~\mathrm{N\,m^{-1}}$). The undamped model~\eqref{eq:lumped_io} instead over-predicts the whole energy budget, because it already over-predicts the exit velocity and the cubic flux roughly triples that excess, so only the radiation-damped model is plotted in figure~\ref{sif_energy}.}

\begin{figure*}[htbp]
	\centering
	\includegraphics[width=0.62\columnwidth]{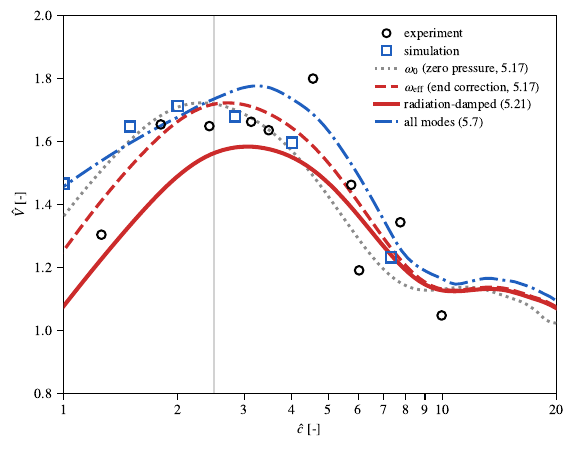}
	\caption{\color{black}Resonance of the peak exit-jet velocity ratio $\hat V=u_{\max}/u_{\max,\mathrm{rigid}}$, read at the first overshoot, versus the nondimensional wave speed $\hat c=cT_{\mathrm{acc}}/L$. The thick solid line is the radiation-damped single-mode model~\eqref{eq:lumped_damped}. The dashed and dotted lines are the undamped closures with ($\omega_{\mathrm{eff}}$, $L_e=R$) and without ($\omega_0$) the end correction. The blue dash-dotted line is the all-mode solve of~\eqref{eq:source} with the same radiation-damped closure, which lies above the single mode over $\hat c\approx3$--$6$ by retaining short-wavelength content the real flow does not sustain (appendix~\ref{app:1dcollapse}). Open squares are the present simulations and open circles the PIV experiments of \citet{choi2026squid}. The faint vertical grey line marks the single-mode-truncation limit $\hat c\approx2.5$, below which the single-mode reduction loses accuracy, defined as the point where the leading dropped mode reaches $\lvert Q_2\rvert/\lvert Q_1\rvert=0.1$ (see figure~\ref{fig:modetrunc} in appendix~\ref{app:1dcollapse}). All curves peak near $\hat c\approx3$, the superpropulsion optimum.}
	\label{sif_squid}
\end{figure*}

\textcolor{black}{Finally, we use each closure to plot the overshoot of the flexible nozzles against the nondimensional wave speed $\hat c=cT_{\mathrm{acc}}/L$, built from the Moens--Korteweg speed $c=\sqrt{Eh/(\rho D)}$ (figure~\ref{sif_squid}), taking the overshoot as the first peak of the exit-jet velocity ratio $\hat V=u_{\max}/u_{\max,\mathrm{rigid}}$. The closures are compared on the figure with the present simulations (open squares) and the soft-nozzle experiments of \citet{choi2026squid} (open circles). Computed with the radiation-damped model~\eqref{eq:lumped_damped}, the ratio peaks at $\hat V=1.58$ near $\hat c\approx3$, where the nozzle recoil is phase-matched to the jet pulse, against $\hat V=1.68$ at the simulated optimum $\hat c\approx2.8$ and $\hat V\approx1.6$--$2.1$ across the experimental optimum band $\hat c\approx2$--$4$. The undamped closures lie above the radiation-damped curve, the zero-pressure $\omega_0$ and the end-corrected $\omega_{\mathrm{eff}}$ both peaking slightly higher at $\hat V=1.72$ for want of damping. The optimum itself also differs slightly between the closures, moving from $\hat c\approx2.3$ for $\omega_0$ to $2.7$ for $\omega_{\mathrm{eff}}$ and $\approx3$ for the radiation-damped model, because the end correction and the ejection-phase radiation loss each lower the recoil frequency at a given stiffness, so a slightly faster wave is required to stay phase-matched to the pulse.} \textcolor{black}{This optimum agrees roughly with the soft-nozzle experiments~\citep{choi2026squid,choi2024mechanism} and the present work, which locate it only within the broad band $\hat c\approx2$--$4$. All three single-mode closures under-predict the simulations at low $\hat c$. At $\hat c=1$ the simulated ratio is $\hat V\approx1.47$, which $\omega_0$, $\omega_{\mathrm{eff}}$ and the radiation-damped model under-predict by $7\%$, $15\%$ and $27\%$ respectively. This deficit is the signature of the single-mode truncation, as at low $\hat c$ the shorter driving wavelength excites higher modes that the leading-mode projection discards. Solving the full flow-rate wave equation~\eqref{eq:source} with the same radiation-damped closure (blue dashed in figure~\ref{sif_squid}) supports this. It recovers the low-$\hat c$ points ($\hat c<2$) but over-predicts over $\hat c\approx3$--$6$, where it retains short-wavelength content the real flow dissipates. The leading mode dominates once the driving wavelength $\lambda/L=2\hat c$ leaves the nozzle acoustically compact, i.e. for $\hat c\gtrsim2.5$, where $\lvert Q_2\rvert/\lvert Q_1\rvert$ falls below $0.1$ (figure~\ref{fig:modetrunc}). All four present cases ($\hat c=2.8$--$32$) lie in this range.}

\textcolor{black}{The reduced model rests on a chain of assumptions, which we collect here. (i) The tube law is linear (Hookean), $\mathrm dp=\mathrm dA/C$, valid for small wall strain. The measured area dilation stays within this range, $\Delta A/A_0\lesssim8\%$ even for the most compliant nozzle (\S\ref{sec:model}). (ii) The flow is inviscid (the wall-shear term is $O(10^{-3})$ and is dropped), so no viscous or structural damping is carried. The observed wall ringing is only lightly damped, and the sole dissipation retained is the vortex-radiation sink of~\eqref{eq:lumped_damped}. (iii) The Galerkin projection keeps a single mode, whose truncation error obeys $\lvert Q_2\rvert/\lvert Q_1\rvert\propto1/(2n-1)^3$, small for $\hat c\gtrsim2.5$ but growing at lower $\hat c$, where the all-mode solve recovers the simulated overshoots that the single mode under-predicts (figure~\ref{sif_squid} and appendix~\ref{app:1dcollapse}). (iv) The inertial end correction is fixed at the rigid, self-similar value $L_e=R$. The self-similar source-disk picture that fixes $L_e$ fails for the compliant walls, where the source spreads onto the expanding side wall and $L_e$ should drift in time (appendix~\ref{app:selfsimilar}). (v) The exit area is held constant in the output quantities $I_m$ and $u_{\mathrm{out}}$. (vi) The exit closure is a quasi-steady radiation ansatz, which reinstates the kinetic energy each slug convects away, $\tfrac12\rho A\,u_{\mathrm{out}}^{3}$, as an equivalent boundary resistance, as for an open pipe end in acoustics \citep{levineschwinger1948,kinsler2000fundamentals}. The pressure closure~\eqref{eq:closure_loss} and the oscillator~\eqref{eq:lumped_damped} also treat the Bernoulli head $\tfrac12\rho u_{\mathrm{out}}^{2}$ differently. The pressure closure retains it, an exact consequence of the momentum theorem, whereas the oscillator omits the reversible head and carries only the radiation over-pressure, a deliberate repair that installs the energy sink, since reinstating the head would cancel the over-pressure on ejection and the oscillator would again conserve energy.}

\textcolor{black}{Despite these limitations,} by collapsing the coupled fluid--structure interaction onto a single compliance-tuned oscillator, the model offers a design tool wherever pulsatile flow drives a deformable conduit. The same one-dimensional compliant-tube wave description underlies arterial pulse-wave models in cardiovascular mechanics \citep{vandevosse2011pulse,hughes1973onedimensional}, where the boundary inertance carried by the end correction maps onto the outflow loading of a vessel segment. In bio-inspired engineering it supplies a closed-form resonance criterion for tuning underwater soft robots and pulsed-jet thrusters to their optimal wave speed \citep{bujard2021resonant,choi2026squid}, complementing vortex-formation principles for biological propulsion \citep{dabiri2009optimal}.

\color{black}
\vspace{-0.4cm}

\section{Conclusion}

We examined stiffness-controlled wave propagation in passively deforming compliant nozzles ($Eh = 75-500~\mathrm{N\,m^{-1}}$) and its influence on vortex dynamics, hydrodynamic impulse, and energy exchange. Under a pulsatile jet input, the nozzle expands during acceleration and contracts during deceleration, generating a travelling wall-deformation wave. Decreasing stiffness reduces the wave speed, consistent with the Moens--Korteweg scaling $c \propto \sqrt{Eh}$, while the damped natural frequency increases monotonically with stiffness. Slower wave propagation in more compliant nozzles extends the expansion phase, increasing entrainment and delaying shear-layer roll-up, thereby weakening vortex growth during the initial acceleration stage. During the transition from peak expansion to contraction, the entrained fluid is expelled rapidly, producing vortex rings with higher circulation and convection speeds, yielding a larger hydrodynamic impulse. An energy budget shows that elastic potential energy stored during expansion is released during contraction, amplifying the outlet kinetic energy flux.
Overall, stiffness-controlled wave timing provides a passive means to tune the pulsed-jet/nozzle interaction and enhance thrust, relevant to underwater propulsion.
\textcolor{black}{To rationalise these observations, we developed a reduced-order model that condenses the coupled fluid--structure response into a lumped store-and-release oscillator. Starting from the one-dimensional mass and momentum conservation equations of a compliant tube, closed by the linear tube law, the inviscid balance collapses to a wave equation for the flow rate, and a single-mode Galerkin projection then yields an input--output ODE. The ODE is closed at the exit by two terms: (i) an inertial end correction, fixed to the geometric value $L_e=R$ by modelling the efflux as a self-similar source disk, and (ii) a vortex-radiation damping that acts only during ejection, its magnitude fixed by the momentum theorem for the discharged jet (the Borda--Carnot limit), whose rate of work matches the kinetic energy each ejected slug carries off with its vortex ring. The end correction sets the resonance $\omega_{\mathrm{eff}}=\omega_0/\sqrt{1+2\epsilon}$ and reproduces the simulated ringing frequency to within $\sim4\%$ on average across $Eh=75$--$500~\mathrm{N\,m^{-1}}$, while the radiation damping reproduces the post-overshoot velocity decay that the undamped closures cannot. Compared against the simulations in dimensional units, the model reasonably follows the total hydrodynamic impulse and its momentum and pressure components, the momentum impulse to within $4\%$ at every stiffness, while the pressure impulse of the compliant nozzles is over-predicted in both timing and peak. The energy budget shows the same store-and-release signature, the outlet kinetic-energy flux recovered to within $14\%$ for the most compliant nozzle ($Eh=75~\mathrm{N\,m^{-1}}$), and the release-phase elastic energy tracked closely where the undamped closure over-predicts it by more than half. The reduction rests on stated assumptions, namely a Hookean small-strain tube law, an inviscid balance whose only retained dissipation is the vortex-radiation sink, a single-mode truncation that loses accuracy at low $\hat c$ where the shorter driving wavelength excites higher modes, a self-similar end correction $L_e=R$ that fails once the compliant wall spreads the exit source, and a quasi-steady ansatz for the exit radiation over-pressure. Within these limits, the model turns stiffness-controlled wave timing into a compact, predictive design principle, whereby matching the nozzle recoil to the jet pulse passively tunes the pulsed-jet/nozzle interaction to maximise thrust, offering a route to optimise bio-inspired underwater propulsors without resolving the full coupled dynamics.}

\section*{Supplementary materials}
Supplementary video 1 demonstrates the elastic wave propagation through the flexible nozzle and the resulting strain contour of the nozzle. Supplementary video 2 shows the generation of vortex rings at the nozzle end and their downstream convection for different flexibilities.

\section*{Funding}
This work was supported by the DARPA Young Faculty Award (grant no.~DARPA-RA-24-01-18-YFA18-FP-004, PI: Saad Bhamla) and the National Research Foundation of Korea (grant no.~RS-2022-NR070924, PI: Daehyun Choi). The content of the information does not necessarily reflect the position or the policy of the Government, and no official endorsement should be inferred. Approved for public release. Distribution is unlimited. The present fluid–structure interaction simulations are carried out using the computational resources provided by the UK national supercomputing facility ARCHER2 through the EPSRC Access To HPC Pioneer Grant (PI: Chandan Bose).

\section*{Declaration of interests}
The authors report no conflict of interest.

\section*{Data availability}
\textcolor{black}{The code, raw data and plot-ready data reproducing the reduced-order model of \S\ref{sec:model} and appendices~\ref{app:analogy}--\ref{app:1dcollapse}, together with figures~\ref{fig:outlet_pressure}--\ref{fig:modetrunc}, are openly available at \url{https://github.com/BioFSILab/FlexibleNozzle-PulsedJet-JFM-2026}.}

\appendix
\color{black}
\renewcommand{\theHsection}{appendix.\thesection}

\section{Viscous derivation, canonical analogies, and the inviscid wave equation}
\label{app:analogy}

The store-and-release enhancement reported above is a particular realisation of superpropulsion, the resonant amplification by which elastic energy stored during acceleration is released in phase with the drive. To expose this mechanism we recover the damped-wave structure that the present jet--nozzle system shares with two canonical dissipative resonators, an RLC transmission line and a spring--damper--mass chain. The main text dropped the small $O(10^{-3})$ wall-shear term to obtain the inviscid wave equation~\eqref{eq:q_wave_nondim} for the leading-order dynamics. Here we instead retain it, since the damping it introduces is what makes the analogy explicit.

Retaining the wall friction, we assume axisymmetric laminar (Poiseuille) flow in a circular, non-permeable tube. The axial velocity profile is parabolic, $u(r)=2\,\bar{u}\,(1-r^{2}/R^{2})$ with $\bar{u}=q/A$, giving the wall shear \textcolor{black}{$\tau_w=\mu(\partial u/\partial r)|_{r=R}=-4\mu\,\bar{u}/R=-(4\mu)/(\pi R^{3})\,q$, negative in the convention fixed in \S\ref{sec:model} because the wall retards the flow. Hence $(1/\rho)\oint_{l}\tau_w\,dl=(2\pi R/\rho)\,\tau_w=-(8\pi\mu)/(\rho A)\,q$, a linear friction with the positive resistance per unit length $\mathcal{R}=8\pi\mu/A$, so the axial momentum balance becomes $\partial q/\partial t+(A/\rho)\,\partial p/\partial x=-(\mathcal{R}/\rho)\,q$, the retarding sign now following from the signed $\tau_w$ rather than being imposed.} Taking $\partial/\partial t$ of the momentum equation and eliminating $p$ with the continuity relation ($C=A/\rho c^2$) yields the damped wave equation

\begin{equation}
\frac{\partial^2 q}{\partial t^2}
 + \frac{8\pi \mu}{\rho A}\,\frac{\partial q}{\partial t}
 = c^2\,\frac{\partial^2 q}{\partial x^2}.
\label{eq:q_wave_laminar}
\end{equation}

\begin{figure*}[t!]
    \centering
    \includegraphics[width=\textwidth]{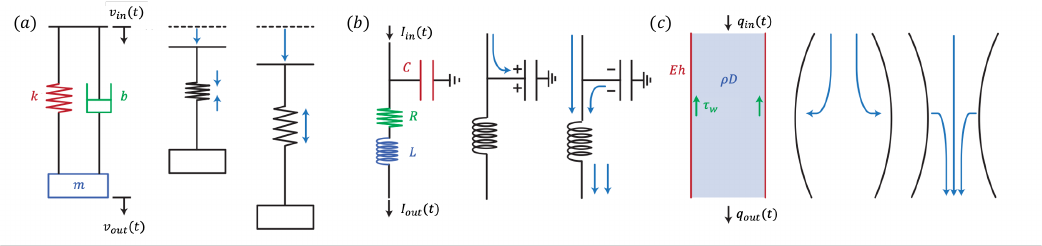}
    \caption{\color{black}\textbf{Superpropulsion analogies.}
    (\textbf{A})~Spring–damper–mass resonator.
    (\textbf{B})~Electrical RLC resonator.
    (\textbf{C})~Current flexible nozzle system.
    On the right, a schematic of superpropulsion as a store–and–release process, in which elastic energy stored during acceleration is subsequently released to amplify the response. Here, $k$ is the spring constant, $b$ is the damping coefficient, $m$ is the mass, $v$ is velocity, $C$ is capacitance, $R$ is resistance, $L$ is inductance, $Eh$ is structural stiffness, $\tau_w$ is wall shear stress, and \textcolor{black}{$k$ in panel (\textit{a}) is an axial stiffness [N], the one-dimensional analogue of $EA$. The grouping $\rho D$ in panel (\textit{c}) is the fluidic mass per unit area that sets the Moens--Korteweg speed $c=\sqrt{Eh/(\rho D)}$, the mass per unit length of the enclosed fluid column being $\rho A$.}}
    \label{sif_mathmodel_lumped}
\end{figure*}

The damped wave equation~\eqref{eq:q_wave_laminar} takes the same form as two canonical distributed systems under a continuum assumption (Fig.~\ref{sif_mathmodel_lumped}). First, an RLC transmission line (per–unit–length constants $\mathcal{L}$ [H/m], $\mathcal{R}$ [$\Omega$/m], $\mathcal{C}$ [F/m]). Combining the telegrapher's equations for the current $I(x,t)$ gives

\begin{equation}
\frac{\partial^{2} I}{\partial t^{2}} + \frac{\mathcal{R}}{\mathcal{L}}\frac{\partial I}{\partial t}
 = \frac{1}{\mathcal{L}\mathcal{C}}\,\frac{\partial^{2} I}{\partial x^{2}}.
\end{equation}

\noindent Secondly, a continuous spring–damper–mass chain (continuum limit of a linear lattice). With mass density $m_\ell$ [kg/m], viscous damping per unit length $c_d$ [N$\cdot$s/m$^{2}$], and \textcolor{black}{axial stiffness $k$ [N], a stiffness multiplied by a length and hence the one-dimensional analogue of $EA$ rather than a stiffness per unit length}, the velocity state $v(x,t)=\partial u/\partial t$ obeys

\begin{equation}
\frac{\partial^{2} v}{\partial t^{2}} + \frac{c_d}{m_\ell}\,\frac{\partial v}{\partial t}
 = \frac{k}{m_\ell}\,\frac{\partial^{2} v}{\partial x^{2}}.
\end{equation}

\noindent Note that we express the spring–damper–mass equation in terms of velocity $v$ rather than displacement $u$, because flow rate $q$ (volume per unit time) and current $I$ (charge per unit time) are both rate quantities. Using velocity $v = \partial u/\partial t$ (displacement per unit time) establishes a direct correspondence among the three systems. All three equations share the same damped wave structure, with an inertia term, a damping term proportional to the first time derivative, and a stiffness term proportional to the spatial second derivative. In these analogies, the flow rate $q$ plays the role of current $I$ or velocity $v$. The squared wave speed $c^{2}$ corresponds to the inverse capacitance-inductance product $1/(\mathcal{L}\mathcal{C})$ or the stiffness-to-mass ratio $k/m_\ell$. The viscous damping coefficient $8\pi\mu/(\rho A)$ corresponds to the resistance-to-inductance ratio $\mathcal{R}/\mathcal{L}$ or the damping-to-mass ratio $c_d/m_\ell$.

The analogy of different systems suggests that the current jet-nozzle system shares the same resonant-matching mechanism, so-called superpropulsion, which has been explored across systems ranging from elastic projectiles \citep{celestini2020contactlayer,giombini2022use}, and in droplet superpropulsion \citep{prl119108001} known as rapid excretion mechanism of sharpshooter as biological example \citep{challita2023superpropulsion}.

The underlying physics of superpropulsion can be understood through a spring-mass system propelled by a harmonically oscillating plate \citep{celestini2020contactlayer}. The actuating plate undergoes unidirectional motion, accelerating the projectile while compressing the spring, which stores elastic energy (Fig.~\ref{sif_mathmodel_lumped}, A). An optimally tuned elastic projectile delays the force release so that the contact force becomes in-phase with the plate velocity, maximizing the instantaneous mechanical power and resulting in the energy transfer factor up to 300\% when the frequency ratio $f_0/f \approx 1.62$ (where $f_0$ and $f$ mean eigenfrequency and driving frequency, respectively) \citep{celestini2020contactlayer}.

\section{Unsteady-Bernoulli derivation of the outlet closure}
\label{app:bernoulli}

Here we derive the outlet over-pressure closure \eqref{eq:bc_integral}. Just outside the exit the efflux must accelerate the surrounding fluid. We follow the centreline streamline from the exit plane to the quiescent far field ($u\to0$, $p\to0$), with $x$ the axial distance measured outward from the exit. For inviscid flow along this streamline the Euler momentum balance is

\begin{equation}
\rho\left(\frac{\partial u}{\partial t}+u\,\frac{\partial u}{\partial x}\right)=-\frac{\partial p}{\partial x} .
\label{eq:app_euler}
\end{equation}

\noindent Using the identity $u\,\partial_x u=\partial_x(u^2/2)$, dividing by $\rho$ and integrating from the exit plane to infinity gives

\begin{equation}
\int_{\mathrm{out}}^{\infty}\frac{\partial u}{\partial t}\,\mathrm{d}x+\Big(\tfrac12 u_\infty^{2}-\tfrac12 u_{\mathrm{out}}^{2}\Big)=-\frac{1}{\rho}\big(p_\infty-p_{\mathrm{out}}\big) ,
\label{eq:app_bernoulli_int}
\end{equation}

\noindent where the two pure-derivative integrals reduce to their endpoint differences. The far-field conditions $u_\infty=0$ and $p_\infty=0$ leave, after rearrangement, exactly the closure \eqref{eq:bc_integral},

\begin{equation*}
\frac{p_{\mathrm{out}}(t)}{\rho}=\int_{\mathrm{out}}^{\infty}\frac{\partial u}{\partial t}\,\mathrm{d}x-\tfrac12 u_{\mathrm{out}}^{2} ,
\end{equation*}

\noindent with $u_{\mathrm{out}}=q_{\mathrm{out}}/A$. The first term is the inertial over-pressure that accelerates the external added-mass column, and the second is the steady Bernoulli suction.

\section{Self-similarity of the rigid-nozzle end correction}
\label{app:selfsimilar}

Here we justify the constant end-correction length $L_e=R$ used in the outlet closure \eqref{eq:bc_endcorrection}, by showing that for a rigid nozzle the exit added-mass column is geometrically self-similar, so that $L_e$ is a time-independent property of the nozzle radius alone.

\begin{figure*}[htbp]
	\centering
	\begin{minipage}[t]{\textwidth}
		\centering
		\includegraphics[width=\linewidth]{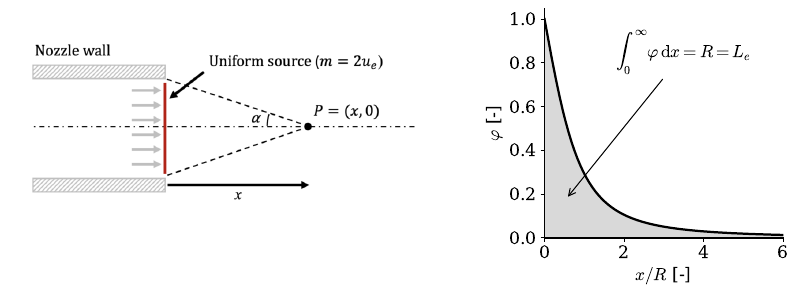}
		\caption{\color{black}The uniform source-disk model behind the end correction. \textit{(a)} The uniform efflux $u_{\mathrm{out}}$ leaving the nozzle is modelled as a permeable uniform source disk of radius $R$ in unbounded fluid. A field point $P=(x,0)$ on the axis sees the disk under a cone of half-angle $\alpha$, giving $\varphi=1-\cos\alpha=1-x/\sqrt{x^2+R^2}$. \textit{(b)} The on-axis profile $\varphi(x)$ and its column integral $\int_0^\infty\varphi\,\mathrm{d}x=R\equiv L_e$.}
		\label{sif_source_disk}
	\end{minipage}

	\vspace{1ex}

	\begin{minipage}[t]{\textwidth}
		\centering
		\includegraphics[width=\linewidth]{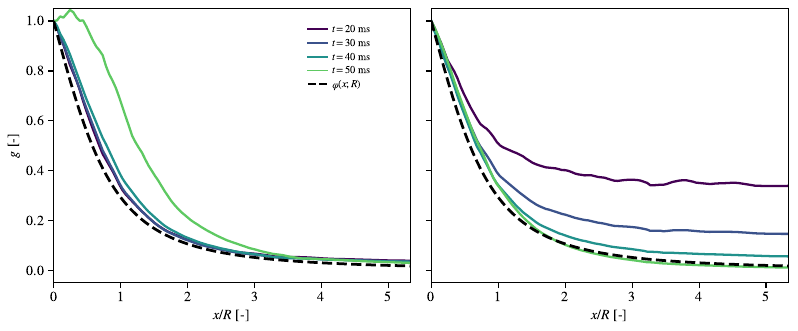}
		\caption{\color{black}Direct numerical-simulation test of self-similarity, showing the normalised acceleration kernel $g(x,t)=\partial_t u(x,t)/\partial_t u(0,t)$ at four rise-phase instants, compared with the source-disk profile $\varphi(x;R)$ (dashed). \textit{(a)} For the rigid nozzle the curves collapse onto $\varphi$, confirming a time-invariant added-mass column and a constant $L_e=R$. \textit{(b)} For the most compliant nozzle ($Eh=75~\mathrm{N\,m^{-1}}$) the wall storage broadens the column early and the curves do not collapse, so $L_e(t)$ drifts during the pulse.}
		\label{sif_selfsimilar}
	\end{minipage}
\end{figure*}

The added-mass over-pressure in \eqref{eq:bc_integral} is set by the external velocity field driven by the efflux. We model the exit plane as a permeable uniform circular source disk of radius $R$ in unbounded fluid, where each surface element emits isotropically, with surface source density chosen so that the on-axis exit velocity equals the bulk efflux $u_{\mathrm{out}}$. This construction, which superposes simple sources over a circular disk, is the classical Rayleigh-integral treatment of a vibrating circular piston \citep[][Vol.~II, \S278]{rayleigh1896theory,lamb1932hydrodynamics}, and the profile derived below is precisely the incompressible ($ka\to0$) limit of the standard on-axis piston near field \citep[][\S7.4]{kinsler2000fundamentals}. Superposing the elementary point sources, a field point on the axis at distance $x$ from the disk subtends a cone of half-angle $\alpha$ with $\cos\alpha=x/\sqrt{x^2+R^2}$, and the induced on-axis velocity is

\begin{equation}
u(x,t)=u_{\mathrm{out}}(t)\,\varphi(x),\qquad
\varphi(x)=1-\cos\alpha=1-\frac{x}{\sqrt{x^{2}+R^{2}}} .
\label{eq:app_phi}
\end{equation}

\noindent The shape $\varphi(x)$ depends only on the geometric ratio $x/R$ and is independent of time. The corresponding end-correction length is its column integral,

\begin{equation}
L_e=\int_0^\infty\varphi(x)\,\mathrm{d}x
=\int_0^\infty\!\left(1-\frac{x}{\sqrt{x^{2}+R^{2}}}\right)\mathrm{d}x
=\Big[\,x-\sqrt{x^{2}+R^{2}}\,\Big]_0^\infty
= R ,
\label{eq:app_Le}
\end{equation}

\noindent since $x-\sqrt{x^2+R^2}\to0$ as $x\to\infty$ and equals $-R$ at $x=0$. Thus the closure length is $L_e=R$, fixed entirely by the nozzle radius (Fig.~\ref{sif_source_disk}). It is instructive to compare this on-axis permeable-efflux length with the end corrections reported for related configurations, all expressed here as an equivalent inertial column length. The impermeable translating disk, whose broadside added mass is $m_a=\tfrac{8}{3}\rho R^{3}$, gives $L_e=8R/(3\pi)\approx0.85R$ \citep{lamb1932hydrodynamics}. The flanged (baffled) acoustic outlet gives $L_e=8R/(3\pi)\approx0.85R$ \citep{rayleigh1896theory}. The unflanged circular pipe gives $L_e\approx0.61R$ \citep{levineschwinger1948}. \textcolor{black}{The present value $L_e=R$ is the largest of these not because the jet is nonlinear, the source-disk field being an entirely linear potential solution, but because it is an on-axis column length whereas the classical corrections are area-averaged reactions over the exit plane. The induced velocity on the axis exceeds its average across the disk, so the same linear field yields the longer length. The distinction is one of the quantity averaged, and it is the on-axis value that the centreline closure~\eqref{eq:bc_integral} requires.}

Equation~\eqref{eq:app_phi} expresses self-similarity, since the acceleration column $\partial_t u(x,t)=\dot u_{\mathrm{out}}(t)\,\varphi(x)$ retains a fixed spatial shape, so the added-mass integral factorises as $\int_0^\infty\partial_t u\,\mathrm{d}x=\dot u_{\mathrm{out}}\!\int_0^\infty\varphi\,\mathrm{d}x=L_e\,\dot u_{\mathrm{out}}$ with $L_e=R$ constant in time. This is the condition under which the end-correction closure \eqref{eq:bc_endcorrection} is exact. We verify it directly from the numerical simulations by forming the normalised acceleration kernel $g(x,t)=\partial_t u(x,t)/\partial_t u(0,t)$. For the rigid nozzle the kernel collapses onto $\varphi(x;R)$ at all rise-phase instants (core RMSE $\approx0.09$, no fitting, figure~\ref{sif_selfsimilar}\textit{a}), and the instantaneous column length $L_e(t)/R=\int g\,\mathrm{d}x/R$ sits in a narrow band ($\approx1.3$--$1.4$) throughout the pulse. The rigid added-mass column is therefore \textcolor{black}{time-invariant, so a single constant $L_e$ is well defined for the rigid nozzle. That measured column, $L_e(t)/R\approx1.3$--$1.4$, sits above the analytic source-disk value $L_e=R$ because the rolled-up shear layer adds an on-axis acceleration that the attached source disk omits, the same deficit that makes the closure under-predict the rigid exit-pressure amplitude in figure~\ref{fig:outlet_pressure}(\textit{a}). We nonetheless carry the parameter-free geometric value $L_e=R$ throughout, so that the closure introduces no quantity taken from the simulations, and the $\sim6\%$ frequency agreement of figure~\ref{fig:resonance} is obtained with that choice.}

For a flexible nozzle the wall expands and stores fluid volume during acceleration and recoils during deceleration. This store-and-release diverts flow away from the exit early in the pulse, broadening the acceleration column ($L_e(t)/R$ rising to $\sim5$ for the most compliant nozzle), and elastic recoil then refocuses it onto the exit ($L_e(t)/R\to1$). The kernel $g(x,t)$ consequently does not collapse (figure~\ref{sif_selfsimilar}\textit{b}), and the instantaneous end correction $L_e(t)$ drifts during the pulse. Treating $L_e$ as the constant rigid value $L_e=R$ is thus a leading-order approximation for the compliant cases, justified because the resulting period prediction $T_{\mathrm{eff}}=T_0\sqrt{1+2\epsilon}$ already matches the simulated wall ringing to within $\sim4\%$ on average (figure~\ref{fig:resonance}). The residual time dependence of $L_e$ is a higher-order correction arising specifically from the time-dependent nozzle expansion absent in the rigid case.

\section{Projection integrals \textcolor{black}{and radiation-damping reduction} for the single-mode balance}
\label{app:projection}

With the single-mode form $\tilde q(x,t)=\tilde q(1,t)\,\phi(x)$ and $\phi(x)=\sin(\pi x/2)$ (so $\phi(0)=0$, $\phi(1)=1$ and $\phi'(x)=\tfrac{\pi}{2}\cos(\pi x/2)$), the projection of the source equation~\eqref{eq:source} onto $\phi$ uses three elementary integrals,

\begin{equation}
\int_0^1\phi^2\,\mathrm{d}x=\tfrac12,\qquad
\int_0^1\phi\,\mathrm{d}x=\tfrac{2}{\pi},\qquad
\int_0^1(\phi')^2\,\mathrm{d}x=\frac{\pi^2}{8},
\end{equation}

\noindent which follow from the half-angle identities $\sin^2\theta=\tfrac12(1-\cos2\theta)$ and $\cos^2\theta=\tfrac12(1+\cos2\theta)$ with $\sin\pi=0$. The three projected terms are then obtained directly. The inertia term is

\begin{equation}
\int_0^1\phi\,\ddot{\tilde q}\,\mathrm{d}x=\ddot{\tilde q}(1,t)\int_0^1\phi^2\,\mathrm{d}x=\tfrac12\,\ddot{\tilde q}(1,t).
\end{equation}

\noindent The stiffness term, integrated by parts once, is

\begin{equation}
\hat c^2\!\int_0^1\phi\,\tilde q''\,\mathrm{d}x
=\hat c^2\Big\{\big[\phi\,\tilde q'\big]_0^1-\int_0^1\phi'\,\tilde q'\,\mathrm{d}x\Big\}
=\hat c^2\,\tilde q'(1,t)-\frac{\pi^2}{8}\,\hat c^2\,\tilde q(1,t),
\end{equation}

\noindent where the boundary bracket reduces to $\phi(1)\,\tilde q'(1,t)=\tilde q'(1,t)$ because $\phi(0)=0$, and $\int_0^1\phi'\,\tilde q'\,\mathrm{d}x=\tilde q(1,t)\int_0^1(\phi')^2\,\mathrm{d}x=\tfrac{\pi^2}{8}\,\tilde q(1,t)$. The surviving boundary term $\hat c^2\,\tilde q'(1,t)$ is left unevaluated and carries the outlet condition weakly, to be closed in the main text. The source term is

\begin{equation}
\int_0^1\phi\,\ddot q_{\mathrm{in}}\,\mathrm{d}x=\ddot q_{\mathrm{in}}\int_0^1\phi\,\mathrm{d}x=\tfrac{2}{\pi}\,\ddot q_{\mathrm{in}}.
\end{equation}

\noindent Collecting the three and multiplying by two gives the single-mode balance~\eqref{eq:Q_open}.

\textcolor{black}{The same projection carries the radiation damping of~\eqref{eq:lumped_damped}, and this second part records that reduction. As assembled in \S\ref{sec:model}, the oscillator carries two terms of the completed closure~\eqref{eq:closure_full}, the inertial end correction and the radiation over-pressure, so the outlet pressure entering the reduction and its rate are}

\textcolor{black}{\begin{equation}
\frac{p_{\mathrm{out}}}{\rho}=L_e\,\dot u_{\mathrm{out}}+\tfrac12\,u_{\mathrm{out}}^{2}H(u_{\mathrm{out}}),\qquad
\frac{\dot p_{\mathrm{out}}}{\rho}=L_e\,\ddot u_{\mathrm{out}}+u_{\mathrm{out}}\dot u_{\mathrm{out}}\,H(u_{\mathrm{out}}),
\label{eq:app_damp_rate}
\end{equation}}

\textcolor{black}{\noindent the derivative of the step dropping out with its factor $u_{\mathrm{out}}^{2}$ (\S\ref{sec:model}). Eliminating $\dot p_{\mathrm{out}}$ with the linearised continuity relation $(A/\rho c^{2})\dot p+q'=0$ and writing $u_{\mathrm{out}}=q_{\mathrm{out}}/A$ turns this rate into the exit slope}

\textcolor{black}{\begin{equation}
q'\big\rvert_{\mathrm{out}}=-\frac{L_e}{c^{2}}\,\ddot q_{\mathrm{out}}-\frac{1}{Ac^{2}}\,q_{\mathrm{out}}H(q_{\mathrm{out}})\,\dot q_{\mathrm{out}},
\label{eq:app_damp_slope}
\end{equation}}

\textcolor{black}{\noindent whose first term is the inertial closure that rescales onto~\eqref{eq:bc_qform} and whose second is its damping companion, the factor $1/A$ born of $u\dot u=q\dot q/A^{2}$. Since the inlet flow is spatially uniform, $q'=\tilde q'$ at the outlet, and the projected balance~\eqref{eq:Q_open} couples to this slope through its weak boundary term, which multiplies it by $2c^{2}/L$ for the damping part just as for the inertial part. The wave speed then cancels between that factor and the $1/c^{2}$ of the slope, so no $c$ survives in the damping, and the balance becomes}

\textcolor{black}{\begin{equation}
(1+2\epsilon)\,\ddot{\tilde q}(1,t)+\frac{2}{LA}\,q_{\mathrm{out}}H(q_{\mathrm{out}})\,\dot q_{\mathrm{out}}+\omega_0^{2}\,\tilde q(1,t)=-\Big(\frac{4}{\pi}+2\epsilon\Big)\ddot q_{\mathrm{in}}.
\label{eq:app_damp_balance}
\end{equation}}

\textcolor{black}{\noindent Dividing by $1+2\epsilon$ (so that $\omega_{\mathrm{eff}}^{2}=\omega_0^{2}/(1+2\epsilon)$), substituting $\tilde q(1,t)=q_{\mathrm{out}}-q_{\mathrm{in}}$ from~\eqref{eq:decomposition} and dropping the same inlet feedthrough dropped between~\eqref{eq:Q_ode} and~\eqref{eq:lumped_io} yields the radiation-damped oscillator~\eqref{eq:lumped_damped} with its prefactor $2/[(1+2\epsilon)LA]$.}

\section{Measurement of the store-and-release period}\label{app:period}

\begin{figure}
	\centering
	\includegraphics[width=0.72\columnwidth]{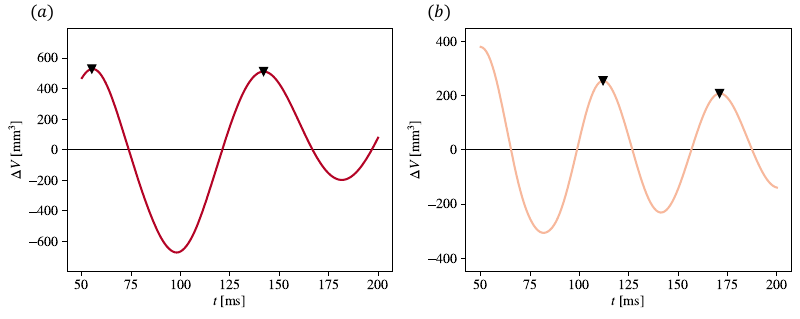}
	\caption{\color{black}Post-pulse wall ringing, showing the detrended stored-volume signal $\Delta V(t)$ with the detected maxima whose mean spacing sets the measured period, for \textit{(a)} the most compliant nozzle ($Eh=75~\mathrm{N\,m^{-1}}$, period $\approx\textcolor{black}{87}$~ms) and \textit{(b)} a stiffer nozzle ($Eh=150~\mathrm{N\,m^{-1}}$, period $\approx59$~ms).}
	\label{sif_ringing}
\end{figure}

The resonance frequency in figure~\ref{fig:resonance} is obtained from the post-pulse ringing of the nozzle wall. We track the stored wall volume $\Delta V(t)=\int_0^L[A(x,t)-A_0]\,\mathrm{d}x$ from the solid solver. Once the inlet pulse has passed its peak at $t=T_{\mathrm{acc}}$, $\Delta V(t)$ oscillates freely about its mean (figure~\ref{sif_ringing}). After removing a slow baseline (\textcolor{black}{a linear trend fitted to the post-pulse record}), the period is measured \textcolor{black}{as the mean spacing of successive maxima}. This yields $31$, $59$, and \textcolor{black}{$87$}~ms for $Eh=500$, $150$, and $75~\mathrm{N\,m^{-1}}$ (frequencies $32$, $17$, and $11$~Hz, i.e. angular frequencies $\omega\approx\textcolor{black}{203}$, $\textcolor{black}{106}$, and $\textcolor{black}{72}~\mathrm{rad\,s^{-1}}$), the measured values plotted in figure~\ref{fig:resonance}. \textcolor{black}{Replacing the linear trend by a constant, a quadratic or a low-pass (Savitzky--Golay) baseline moves the measured periods by at most $1$~ms ($\lesssim2\%$) at every stiffness. A cubic is over-flexible for the softest case, whose post-pulse record holds under two ringing cycles, and begins to absorb the oscillation itself.} \textcolor{black}{Figure~\ref{sif_ringing} shows only $t\ge T_{\mathrm{acc}}$, the window that is detrended and measured. Its ordinate is the residual about the removed trend, not $\Delta V$ itself, so its zero is that trend rather than the undeformed nozzle, and the trace does not begin at zero. The first expansion peak occurs during the driving pulse, so by the time free ringing sets in the wall is already near maximum expansion.}

\section{All-mode one-dimensional model: momentum-impulse collapse}\label{app:1dcollapse}

\textcolor{black}{To isolate what the single-mode reduction costs, we solve the all-mode counterpart of the lumped model: the flow-rate wave equation~\eqref{eq:source} taken just before the Galerkin projection, retaining every mode. Written for $q$ itself the same equation carries a Dirichlet inlet, $q_{tt}=c^{2}q_{xx}$ with $q(0,t)=q_{\mathrm{in}}(t)$. It is integrated by an explicit leapfrog scheme (grid-converged to four digits by $M=80$ nodes) and closed at the exit by the same condition as the lumped model, either the end correction alone or the radiation-damped closure of~\eqref{eq:lumped_damped}. Both models are therefore identical in physics and differ only in the number of retained modes.}

\textcolor{black}{The tube $x\in[0,L]$ is divided into $M$ uniform intervals of width $\Delta x=L/M$, with $q_j^{n}\approx q(j\Delta x,\,n\Delta t)$, and $q_{xx}$ is taken in second-order central form. Away from the ends the explicit leapfrog update is}

\textcolor{black}{\begin{equation}
q_j^{n+1}=2q_j^{n}-q_j^{n-1}+r^{2}\big(q_{j-1}^{n}-2q_j^{n}+q_{j+1}^{n}\big),
\qquad r=\frac{c\,\Delta t}{\Delta x},
\label{eq:leapfrog}
\end{equation}}
\textcolor{black}{\noindent second-order accurate in space and time and stable for the Courant number $r\le1$. We use $r=0.9$. The inlet is imposed strongly, $q_0^{n}=q_{\mathrm{in}}(t^{n})$, and the march starts from rest with the half-step $q^{1}=q^{0}+\tfrac12 r^{2}\delta_x^{2}q^{0}$ that preserves second-order accuracy.}

\textcolor{black}{At the exit node $j=M$ the stencil of~\eqref{eq:leapfrog} reaches the value $q_{M+1}$ at a point outside the tube, a ghost node, and the outlet closure supplies the missing relation that determines it. Pushing the exit pressure through the linearised continuity relation $(A/\rho c^{2})\dot p+q'=0$ writes that closure as a condition on the slope at $x=L$,}

\textcolor{black}{\begin{equation}
\frac{\partial q}{\partial x}\bigg|_{L}=-\frac{L_e}{c^{2}}\,\ddot q\big|_{L}-\frac{G}{c^{2}}\,\dot q\big|_{L},
\qquad G=K_{\mathrm{out}}\max(u_{\mathrm{out}},0)+K_{\mathrm{in}}\max(-u_{\mathrm{out}},0),
\label{eq:outlet_slope_num}
\end{equation}}
\textcolor{black}{\noindent the inertial end correction alone when $G=0$ and the radiation-damped closure otherwise. As in the lumped model we set $K_{\mathrm{out}}=1$ and $K_{\mathrm{in}}=0$, the intake correction being doubly small~\eqref{eq:closure_full}, so $G=\max(u_{\mathrm{out}},0)$ in every solve below. Discretising the slope at step $n$ by the central difference $(q_{M+1}^{\,n}-q_{M-1}^{\,n})/2\Delta x$ turns~\eqref{eq:outlet_slope_num} into an explicit expression for the ghost value, $q_{M+1}^{\,n}=q_{M-1}^{\,n}-(2\Delta x/c^{2})\big(L_e\,\ddot q_M^{\,n}+G\,\dot q_M^{\,n}\big)$, so the ghost node carries no information of its own. Substituting it into the second difference $c^{2}(q_{M-1}^{\,n}-2q_M^{\,n}+q_{M+1}^{\,n})/\Delta x^{2}$ of the wave equation at $j=M$ and collecting the $\ddot q_M^{\,n}$ terms on the left leaves}

\textcolor{black}{\begin{equation}
\Big(1+\frac{2L_e}{\Delta x}\Big)\ddot q_M^{\,n}
=\frac{2c^{2}}{\Delta x^{2}}\big(q_{M-1}^{\,n}-q_M^{\,n}\big)-\frac{2G}{\Delta x}\,\dot q_M^{\,n} .
\label{eq:outlet_node}
\end{equation}}
\textcolor{black}{\noindent Discretising $\ddot q_M^{\,n}$ as in~\eqref{eq:leapfrog} and centring the damping term as $\dot q_M^{\,n}=(q_M^{n+1}-q_M^{n-1})/2\Delta t$ keeps the update explicit, because $G$ is evaluated at step $n$:}

\textcolor{black}{\begin{equation}
q_M^{n+1}=\frac{2q_M^{n}-(1-b)\,q_M^{n-1}
+\dfrac{2r^{2}}{1+2L_e/\Delta x}\big(q_{M-1}^{n}-q_M^{n}\big)}{1+b},
\qquad b=\frac{G\,\Delta t}{\Delta x\,\big(1+2L_e/\Delta x\big)} .
\label{eq:outlet_update}
\end{equation}}
\textcolor{black}{\noindent Setting $L_e=0$ and $G=0$ recovers the zero-pressure outlet $q_x(L,t)=0$. The exit velocity $u_{\mathrm{out}}=q_M/A$ is interpolated onto the $1~\mathrm{ms}$ output grid and post-processed exactly as for the lumped model. The solution is grid converged: refining $M$ from 40 to 320 leaves the peak exit velocity unchanged in the fourth significant figure ($0.5159~\mathrm{m\,s^{-1}}$ at every $M$ for $Eh=75~\mathrm{N\,m^{-1}}$), and the undamped solve agrees to better than $0.4\%$ with an implicit Radau integration of the same semi-discrete system and with the eigenmode sum~\eqref{eq:galerkin} carried to convergence. All results below use $M=80$.}

\begin{figure*}[htbp]
	\centering
	\includegraphics[width=0.66\columnwidth]{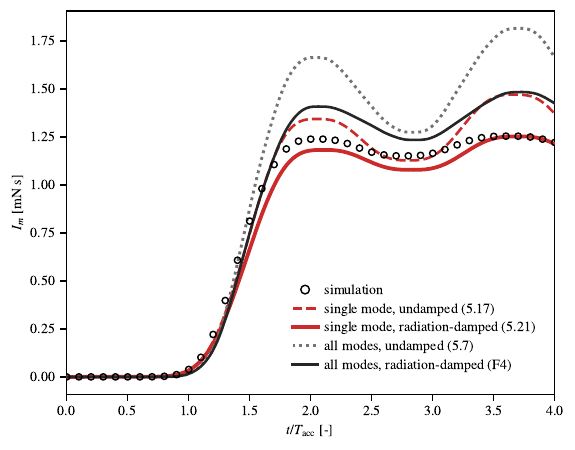}
	\caption{\color{black}Single-mode reduction versus the all-mode solution of eq~\eqref{eq:source}, for the most compliant nozzle ($Eh=75~\mathrm{N\,m^{-1}}$), in dimensional variables, showing the signed momentum impulse $I_m=\rho A\int u_{\mathrm{out}}\lvert u_{\mathrm{out}}\rvert\,\mathrm{d}t$. Circles denote the numerical simulation, with the exit velocity taken as the mass-conserving area average over the exit plane (as in figure~\ref{fig:velocity_history}). Red lines are the single-mode model, dashed undamped~\eqref{eq:lumped_io} and thick solid radiation-damped~\eqref{eq:lumped_damped}. Grey dotted and black solid are the corresponding all-mode solves of~\eqref{eq:source}, the damped one advanced with the outlet update~\eqref{eq:outlet_update}. Retaining every mode raises the impulse away from the measurement rather than towards it.}
	\label{fig:app1dcollapse}
\end{figure*}

\textcolor{black}{Figure~\ref{fig:app1dcollapse} compares them for the most compliant nozzle ($Eh=75~\mathrm{N\,m^{-1}}$) in dimensional variables. Retaining the higher modes does not close the gap to the simulations: averaged over the ringing window $t/T_{\mathrm{acc}}=2$--$4$ the measured momentum impulse is $1.21~\mathrm{mN\,s}$, which the radiation-damped single mode reproduces to within $4\%$ ($1.17$) while the all-mode solve over-predicts it by $13\%$ ($1.36$). Without damping the two read $1.30$ and $1.55$. The higher modes add outlet-flux amplitude that the simulation does not sustain, because in the real flow those short-wavelength components are removed by viscous, three-dimensional and vortex-formation losses that the inviscid wave equation does not contain. Truncating at one mode happens to act as the corresponding filter. The single-mode reduction is therefore not merely a convenience. On the momentum impulse the single-mode model is more accurate than the all-mode model, and the residual deficit of figure~\ref{sif_impulse}(\textit{c}) is not attributable to the Galerkin truncation.}

\begin{figure}
	\centering
	\includegraphics[width=0.66\columnwidth]{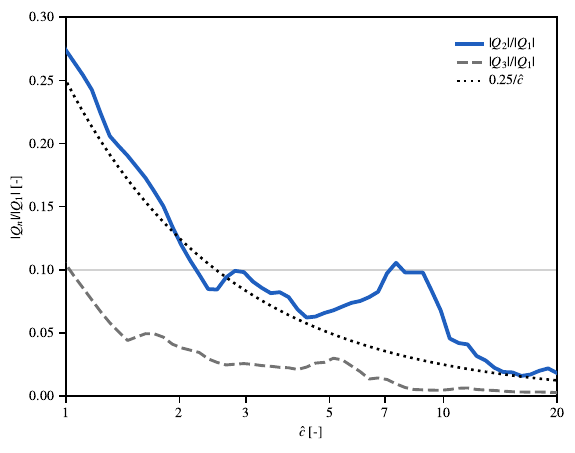}
	\caption{\color{black}Single-mode-truncation error versus the nondimensional wave speed $\hat c$: the modal-amplitude ratios $\lvert Q_2\rvert/\lvert Q_1\rvert$ (solid) and $\lvert Q_3\rvert/\lvert Q_1\rvert$ (dashed), the leading and next dropped modes, computed from the modal oscillators driven by the measured inlet. Both fall as $\hat c$ rises (longer driving wavelength $\lambda/L=2\hat c$, fewer active modes). The faint grey line marks the single-mode threshold $\lvert Q_2\rvert/\lvert Q_1\rvert=0.1$, which the $0.25/\hat c$ law reaches at $\hat c=2.5$. The present cases ($\hat c\ge2.8$) lie below it.}
	\label{fig:modetrunc}
\end{figure}

\textcolor{black}{The range of validity of the single-mode reduction follows from the modal spectrum itself. Projecting~\eqref{eq:source} on each eigenfunction gives independent oscillators $\ddot Q_n+\omega_n^2 Q_n=-c_n\,\ddot q_{\mathrm{in}}$, with the odd-harmonic ladder $\omega_n=(2n-1)\omega_1$~\eqref{eq:eigenfunctions} and modal forcing $c_n=\big(\int_0^1\phi_n\,\mathrm{d}x\big)/\big(\int_0^1\phi_n^2\,\mathrm{d}x\big)=4/[(2n-1)\pi]$. Evaluated at the exit, the expansion~\eqref{eq:galerkin} gives the outlet flux $q(1,t)=q_{\mathrm{in}}(t)+\sum_n Q_n(t)\,\phi_n(1)$, so mode $n$ contributes $Q_n\phi_n(1)$ there, its amplitude weighted by the exit value of its shape. Every eigenfunction ends at the exit at full amplitude, $\phi_n(1)=\sin[(2n-1)\pi/2]=\pm1$, so no mode is attenuated by its shape and dropping mode $n$ changes the outlet flux by exactly $\lvert Q_n\rvert$. The ratio $\lvert Q_2\rvert/\lvert Q_1\rvert$ is therefore the leading truncation error. Two limits bracket this ratio. In the quasi-static limit, where the drive lies well below every $\omega_n$, each oscillator responds through its stiffness, $Q_n\simeq-c_n\ddot q_{\mathrm{in}}/\omega_n^2$. Retaining the forcing projection $c_n\propto1/(2n-1)$ together with $\omega_n=(2n-1)\omega_1$ gives $\lvert Q_n\rvert\propto1/(2n-1)^3$, so the second mode falls to $1/27\approx0.04$ of the first, a stronger suppression than the $1/(2n-1)^2\!\to\!1/9$ that the $1/\omega_n^2$ stiffness scaling alone would suggest. The present nozzles, however, operate near the fundamental resonance ($\hat c\approx2.7$ for the softest), where this estimate fails, since a resonantly driven $Q_1$ is limited by its damping, here the vortex-radiation loss of~\eqref{eq:lumped_damped}, rather than by its stiffness. The justification near resonance comes instead from the frequency content of the drive. Its measured spectrum stays well below $\omega_2=3\omega_1$, so the second mode is driven far from its own resonance and stays small. Evaluated with the measured drive the ratio therefore sits between these bounds, $\lvert Q_2\rvert/\lvert Q_1\rvert\approx0.25/\hat c$ (figure~\ref{fig:modetrunc}), i.e.\ $\approx0.04$--$0.1$ across the present cases, falling below $0.1$ for $\hat c\gtrsim2.5$. The single mode is accurate wherever the driving wavelength $\lambda/L=2\hat c$ is long enough to leave the nozzle acoustically compact, and degrades at small $\hat c$, where the shorter wavelength excites the higher modes. This is a reduced-frequency (spectral) criterion in $\hat c$, distinct from the amplitude criterion $M=V/c$ of \S\ref{sec:model}. All four present cases ($\hat c=2.8$--$32$) lie in the range where $\lvert Q_2\rvert/\lvert Q_1\rvert\lesssim0.1$. The reduction is thus justified not by asymptotic smallness alone (mode 2 is at most $\sim10\%$ near resonance, and $\sim1/27$ once the drive is quasi-static) but by the all-mode test of figure~\ref{fig:app1dcollapse}, which shows that retaining the higher modes over-predicts the outlet flux and worsens the prediction.}

\bibliographystyle{jfm}
\bibliography{jfm}

\end{document}